\documentclass{jfm}
\usepackage[dvips]{graphicx}

\title[DNS of bifurcations in air-filled baroclinic annulus]
{DNS of bifurcations in an air-filled rotating baroclinic annulus}
\author[Randriamampianina, Fr\"uh, Read and Maubert]
 {Anthony Randriamampianina$^1$, Wolf-Gerrit Fr\"uh$^2$, Peter L. Read$^3$,  Pierre Maubert$^1$}
 \affiliation{ $^1$ Institut de Recherche sur les Ph\'enom\`enes Hors  Equilibrie, UMR 6594 CNRS,
 Technop\^ole de Ch\^ateau-Gombert, 49, rue Fr\'ed\'eric Jolie-Curie, BP 146, 13384 Marseille cedex 13, France%
\\
$^2$School of Engineering and Physical Sciences, Heriot-Watt University, Riccarton, Edinburgh, EH14 4AS, UK%
\\
$^3$Atmospheric, Oceanic \& Planetary Physics, University of Oxford, Clarendon Laboratory, Oxford, OX1 3PU, UK%
}

\begin{document}
\maketitle

\begin{abstract}

Three-dimensional Direct Numerical Simulation (DNS) on the nonlinear dynamics and a route to
chaos in a rotating fluid subjected to lateral heating is presented here and discussed in the
context of laboratory experiments in the baroclinic annulus. Following two previous
preliminary studies by Maubert and Randriamampianina (2002, 2003), the fluid used is air
rather than a liquid as used in all other previous work. This study investigated a bifurcation
sequence from the axisymmetric flow to a number of complex flows.

The transition sequence, on increase of the rotation rate, from the axisymmetric solution via
a steady, fully-developed baroclinic wave to chaotic flow followed a variant of the classical
quasi-periodic bifurcation route, starting with a subcritical Hopf and associated saddle-node
bifurcation. This was followed by a sequence of two supercritical Hopf-type bifurcations,
first to an amplitude vacillation, then to a three-frequency quasi-periodic modulated
amplitude vacillation (MAV), and finally to a chaotic MAV\@. In the context of the baroclinic
annulus this sequence is unusual as the vacillation is usually found on decrease of the
rotation rate from the steady wave flow.

Further transitions of a steady wave with a higher wave number pointed to the possibility that
a barotropic instability of the side wall boundary layers and the subsequent breakdown of
these barotropic vortices may play a role in the transition to structural vacillation and,
ultimately, geostrophic turbulence.

\end{abstract}

\section{Introduction and background}
Baroclinic instability is one of the dominant energetic processes in the large-scale
atmospheres of the Earth, e.g.~\cite{PS95}, and other terrestrial planets, such as Mars, and
in the oceans. Its fully-developed form as sloping convection is strongly nonlinear and has a
major role in the transport of heat and momentum in the atmosphere. Its time-dependent
behaviour also exerts a dominant influence on the intrinsic predictability of the atmosphere
and the degree of chaotic variability in its large-scale meteorology,
e.g~\cite{PS95,RCFLL98,R01}. An understanding of the modes of chaotic behaviour associated
with baroclinic processes is therefore of great importance for understanding what determines
the predictability of weather and climate systems.

For more than 50 years, the differentially-heated, rotating cylindrical annulus has been an
archetypal means of studying the properties of fully-developed baroclinic instability in the
laboratory. Laboratory measurements have enabled the transport properties of the flow, its
detailed structure, and aspects of its time-dependent behaviour to be studied under a variety
of conditions, e.g.~\cite{HM75,P80,BPK84,Hetal85,RBJS92,FR97}. The system is well known to
exhibit a rich variety of different flow regimes, depending upon the imposed conditions
(primarily the temperature contrast $\Delta T$ and rotation rate $\Omega$), ranging from
steady, axisymmetric circulations through highly symmetric, regular wave flows to
fully-developed geostrophic turbulence.

Almost all laboratory experiments with the baroclinic annulus and numerical simulations have
used water, water-glycerol mixtures, or silicone oils.  All these have Prandtl numbers much
larger than unity.  One study, by \cite{FP76}, included mercury with $Pr=0.025$ and confirmed
that the observed basic transitions were unchanged even at very small Prandtl numbers.  That
early study, however, did not report on the details of flow transitions within the
non-axisymmetric flow regimes.  Even though air is a very common fluid--and relevant to
atmospheric flows as well as the flow in turbomachinery, no previous detailed study has used
air as the working fluid in the annulus.  Furthermore, the Prandtl number of air is near unity
which could result in revealing phenomena as the relative thickness of the thermal and
velocity boundary layers is similar.  This paper continues two previous studies by Maubert \&
Randriamampianina (2002, 2003) which explored the basic flow types found by direct numerical
simulation of baroclinic convection in an air-filled annulus.

\subsection{Aim and outline}
The focus of this paper is to provide a detailed analysis of the onset of baroclinic waves in
an air-filled rotating annulus, and to follow this through a complete bifurcation sequence up
to a complex fluctuating type of flow, which has not been achieved before with direct
numerical simulation in an even remotely realistic domain such as the simple annulus.  The
findings from the direct numerical simulations are analysed in the framework of standard
bifurcations and discussed in the context of the extensive literature of annulus experiments
with liquids.

The remainder of this section is devoted to a brief background to established bifurcations in
the baroclinic annulus and the role of the Prandtl number.  The model used for the direct
numerical simulations is described in some detail in section 2, followed by a description of
the main techniques used to analyse data from the simulations. Section 4 presents the main
results, including an overview of the main regimes observed and detailed discussions of the
flows encountered (a) from $m = 2$ towards quasi-periodic and chaotic amplitude vacillations,
and (b) from $m = 3$ towards either quasi-periodic amplitude vacillation or structural
vacillation via a mixed vacillation regime not previously identified in other work. The
overall results are discussed in the light of previous work and wider issues in section 5.

\subsection{Bifurcations to chaos in the baroclinic annulus}
\cite{Rand82} has shown that the bifurcations of systems with axially-symmetric geometry and
boundary conditions would be expected to lead to the eventual transition to chaotic behaviour
via a quasi-periodic route, provided the frequencies of various forms of modulation of the
waves remain incommensurate, otherwise a period-doubling route is anticipated. The rotating
annulus seems to form an important prototype for flow systems characterised by axially
symmetric boundary conditions, allowing experimental investigations of the effects of such
imposed symmetries on the bifurcations of the system. Thus, the onset of regular
azimuthally-travelling waves with steady amplitude is commonly found to be the outcome of the
onset of baroclinic instability as parameters such as $\Omega$ are increased from the steady
axisymmetric regime.  In the following, these waves are referred to as \emph{'steady waves'}
and the frequency associated with the azimuthal phase velocity is by convention referred to as
the \emph{'drift frequency'}, $\omega_d$. Secondary bifurcations are found to involve the
onset of periodic modulations of the initial steady, regular waves, leading to various forms
of periodic `vacillation' with a vacillation frequency, $\omega_v$. Further bifurcations may
involve a modulation at a second frequency, $\omega_m$, in the form of a 'modulated
vacillation' or the emergence of other complex flows.  These may follow low-dimensional
chaotic dynamics or signal the emergence of turbulent flow, described as 'weak' or 'strong'
turbulence, respectively by \cite{Rand82}. An important consequence of the axial symmetry of
the boundary conditions, however, is that frequency entrainment of oscillatory amplitude
modulations with the drift frequency of the azimuthal progression of the baroclinic wave
patterns is excluded, on the grounds that there is no physical mechanism to couple these modes
of oscillation.  This complete decoupling of the drift frequency of a travelling wave from any
amplitude variation gives rise to the possibility of observing a quasi-periodic flow with
three free frequencies while a generic three-frequency flow would be likely to be chaotic
(cf.~\cite{NRT78}).  In laboratory experiments, however, such a quasi-periodic flow has never
been observed as even the smallest imperfections are sufficient to provide the necessary
coupling to result in either chaotic flow or frequency entrainment with its resulting
reduction to two independent frequencies.  To our knowledge, also no numerical evidence for a
quasi-periodic three-frequency flow from a numerical simulation of the Navier-Stokes equations
has ever been reported.

Extensive experimental studies of transitions to chaotic behaviour in the laboratory by
\cite{RBJS92} have demonstrated the onset of `baroclinic chaos' via a quasi-periodic route, in
which a periodic amplitude-modulated wave (`amplitude vacillation') develops a long period
secondary modulation, typically involving the instability of a non-harmonic azimuthal sideband
mode. The onset of this secondary modulation was typically found to be chaotic (from
computation of the largest Lyapunov exponent), except when the modulation frequency was
commensurate with the spatial drift frequency. As mentioned above, this apparent entrainment
of the modulation and drift frequencies is not generic for systems with azimuthally-symmetric
boundary conditions, but was almost certainly a consequence of imperfections in the apparatus
and thermocouple array used to measure the azimuthal thermal structure of the flow, breaking
the rotational symmetry of the boundary conditions. Subsequent work by \cite{FR97} has
highlighted that such entrainments with the azimuthal drift frequency were commonplace in the
experiments, indicating that only very small departures from azimuthal symmetry in the
boundary conditions were sufficient to couple intrinsic nonlinear oscillations of the flow
with the azimuthal pattern drift.

\subsection{The role of the Prandtl number}
The role of fluid properties in influencing the nonlinear development of baroclinic flows is
another aspect of the problem which has hitherto received relatively little attention. The
Prandtl number $Pr$ is a parameter of particular interest, also in the context of other
convection problems. \cite{FP76}, who carried out a careful survey of the main flow regimes in
a thermally-driven annulus using either mercury, water, or silicon oils, found significant
differences in the onset of baroclinic instability in the region of the so-called `lower
symmetric transition' at low Taylor number, where viscous and thermal diffusion are expected
to play a major role. Some substantial differences in the onset of various types of regular
wave were also noted at higher Taylor numbers. \cite{J81} investigated the influence of
Prandtl number on the incidence of various forms of vacillation, using fluids with $Pr$
ranging from 11 to 74\@. He reported that amplitude vacillation in particular was
significantly more widespread at high Prandtl number, though the onset of `structural
vacillation' close to the transition zone at high Taylor number was less sensitive to $Pr$. In
all of the published studies so far, however, the range of $Pr$ investigated has either been
limited to relatively high values (using liquids based on water, silicon oils or organic
fluids such as diethyl ether) or very low Pr in liquid metals (mercury). It is remarkable that
no experimental work to date has investigated the properties of baroclinic instability in a
fluid with $Pr$ = O(1).

Air is a commonplace fluid, with a Prandtl number at room temperature of around 0.7, and would
seem to form an ideal medium in which to investigate the properties of fully-developed
baroclinic instability in a fluid with Pr = O(1). Some initial investigations were carried out
recently by \cite{MaRa03} in an air-filled rotating annulus using a pseudo-spectral numerical
model, which have successfully demonstrated the onset of baroclinic instability and the
development of some simple vacillations. That study has now been substantially extended to
investigate the onset of various types of flow in much more detail. The results have exposed a
number of new and intriguing features of fully-developed baroclinic instability in a
numerically simulated environment which is free of experimental imperfections due to probes
and/or departures from circular symmetry, and shed light on a number of issues raised in
previous work.

\section{The numerical model}
\subsection{Governing equations}
\begin{figure}
\centering
\includegraphics[width=0.6\textwidth]{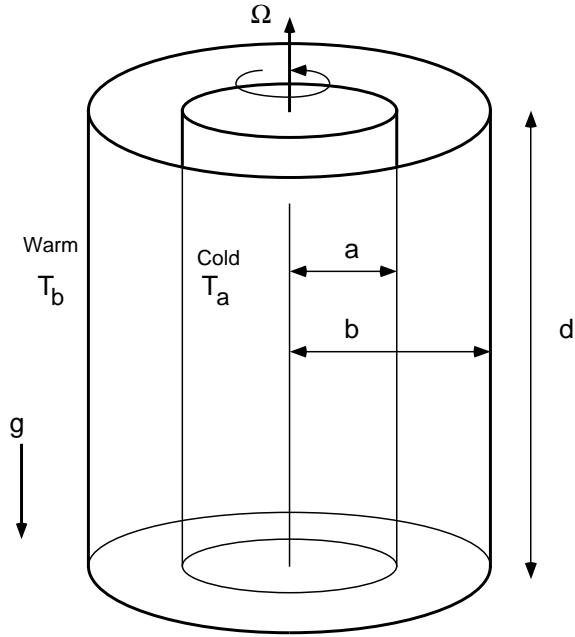}
\caption{\label{Schematic}Schematic diagram of the baroclinic annulus.  }
\end{figure}
\begin{table}
\centering
\begin{tabular}{llrl}
Inner radius & $a$ & 34.8 & mm  \\
Outer radius & $b$ & 60.2 & mm  \\
Height & $d$ & 100 & mm \\
Gap width & $L=b-a$ & 25.4 & mm\\ \hline
Mean temperature & $T_0$ & 293 & $\rm K$\\
Temperature difference & $\Delta T$ & 30 & $\rm K$\\
Rotation rate & $\Omega$ & 10 to 47 & $\rm rad\,s^{-1}$ \\ \hline
Density & $\rho$ & 1.220 & $\rm kg m^{-3}$\\
Volume expansion coefficient & $\alpha=\frac{1}{T_0}$ & $3.41\times10^{-3}$ & $\rm K^{-1}$ \\
Kinematic viscosity & $\nu$ & $1.697\times10^{-5}$ & $\rm m^2 s^{-1}$ \\
Thermal diffusivity & $\kappa$ & $2.400\times10^{-5}$ &$\rm m^2 s^{-1}$ \\ \hline
Aspect ratio & $A=d/L$ & 3.94 & -- \\
Curvature parameter & $R_c=(b+a)/L$ & 3.74 & -- \\
Prandtl number & $Pr=\nu/\kappa$& 0.707 & --\\
Rayleigh number & $Ra=g\alpha(T_b-T_a)L^3 / (\nu \kappa)$ & $4.7332\times10^4$ & --\\
Froude number & $Fr = \Omega^2 L / g$ & 0.313 to 5.730& -- \\
Taylor number & $Ta=4\Omega^2\nu^{-2} L^5 d^{-1}$ & $1.75\times 10^5$ to $32.5\times 10^5$& -- \\
\end{tabular}
\caption{Summary of the dimensions of the system, the fluid properties, and the dimensionless
parameters for the temperature difference of $\Delta T=30\rm K$.} \label{Design}
\end{table}
The physical model is composed of two vertical coaxial cylinders of height $d$, the inner
cylinder of radius $a$ and the outer of radius $b$, resulting in an annular gap of width
$L=b-a$. At the top and the bottom, the annular gap is closed by two horizontal insulating
rigid endplates. The inner cylinder at radius $r = a$ is cooled to a constant temperature,
$T_a$, and the outer cylinder is heated to a constant temperature, $T_b > T_a$, as shown in
Figure~\ref{Schematic}.  The physical dimensions and fluid properties are listed in
Table~\ref{Design}. The whole cavity rotates at a uniform rate, where the rotation vector
${\mathbf \Omega} = \Omega {\mathbf e}_z$ is anti-parallel to the gravity vector ${\mathbf g}
= - g {\mathbf e}_z$ and coincides with the axis of cylinders. The geometry is defined by the
aspect ratio, $A=d/L= 3.94$, and the curvature parameter, $R_c=(b+a)/L=3.74$ (see
Table~\ref{Design}), corresponding to the configuration used by \cite{FH65} in their
experiments with liquids. The working fluid considered in the present study is air at ambient
temperature, so that $Pr=0.707$. The fluid is assumed to satisfy the Boussinesq approximation,
with constant properties except for the density when applied to the Coriolis, centrifugal and
gravitational accelerations where $\rho = \rho_0(1-\alpha(T-T_0))$, where $\alpha$ is the
coefficient of thermal expansion and $T_0$ is a reference temperature $T_0=(T_b+T_a)/2$. The
reference scales are the velocity $U= g \alpha (T_b-T_a)/2 \Omega$ and the time $t=
(2\Omega)^{-1}$, and the nondimensional temperature is $2(T-T_0)/(T_b-T_0)$.  With a mean
temperature of 293K, the Boussinesq approximation can assumed to be valid to temperature
differences of up to 30K.

In the meridional plane, the dimensional space variables $(r^*,z^*)$ $\in$ $[a,b] \times
[0,d]$ have been normalised into the square $[-1,1] \times [-1,1]$, a prerequisite for the use
of Chebyshev polynomials:
\[
r = \frac{2r^*}{L} - R_c, \hspace{4em}   z = \frac{2z^*}{d} -1.
\]

In a rotating frame of reference, the resulting dimensionless system becomes :
\begin{eqnarray} \label {NSE1}
\frac {{\partial{\mathbf V}}}{\partial {t}} +
 \frac{2Ra}{A^2PrTa}N({\mathbf V})+
 {\mathbf e}_z\times{\mathbf V}=-
  \nabla \Pi + \frac{4}{A^{3/2}Ta^{1/2}}\nabla^2 {\mathbf V} +
{\mathbf F}
\end{eqnarray}
\begin{eqnarray} \label {NSE2}
\nabla .{\mathbf V}=0
\end{eqnarray}
\begin{eqnarray} \label {NSE3}
\frac {{\partial T}}{{\partial t}} + \frac {2Ra}{A^2PrTa}\nabla . ({\mathbf V} T) =
\frac{4}{A^{3/2}PrTa^{1/2}}\Delta T
\end{eqnarray}
where  $\mathbf{e}_r$ and $\mathbf{e}_z$ are the unit vectors in the radial and
axial directions respectively,
\begin{eqnarray}
\Pi & = & \frac{p + \rho_0 g z - \frac{1}{2}\rho_0 \Omega^2 r^2}
               {\rho_0 g \alpha (T_b-T_a)d/2} \\
{\mathbf F} & = & \frac{1}{2}T{{\mathbf e}_z} - \frac{Fr}{4A} \left(r + R_c\right) T
{{\mathbf e}_r},
\end{eqnarray}
and $N({\mathbf V})$ corresponds to non-linear terms.  The dimensionless parameters governing
the flow are the aspect ratio $A$, the curvature parameter $R_c$, Prandtl number $Pr$, the
Rayleigh number $Ra$, the Froude number $Fr$, and the Taylor number,
\begin{equation}
    Ta=\frac{4\Omega^2 L^5 }{\nu^2 d}.
\end{equation}
The Taylor number is one of the two main parameters traditionally used to analyse this system,
following e.g.~\cite{FH65} and \cite{HM75}.  The second parameter is the thermal Rossby
number,
\begin{equation}
\Theta= \frac{gd\alpha(T_b-T_a)}{\Omega^2L^2}\equiv \frac{4 Ra}{Pr Ta},
\end{equation}
which explicitly appears as coefficient of the convective terms in (\ref{NSE1}) and
(\ref{NSE3}).

The 'skew-symmetric' form proposed by \cite{zang90} was chosen for the convective
terms $N({\mathbf V})$ in the momentum equations to ensure the conservation of
kinetic energy, a necessary condition for a simulation to be numerically stable in
time.

\subsection{Boundary conditions}
The boundary conditions are no-slip velocity conditions at all surfaces,
$${\bf V} = {\bf 0}\hspace*{2em}\mbox{ at } r=\pm 1 \mbox{ and at }z=\pm1,$$
thermal insulation at the horizontal surfaces,
$$\frac{\partial T}{\partial z} = 0\hspace*{2em}\mbox{ at } z=\pm 1,$$
and constant temperature conditions at the sidewalls,
$$T= -1 \hspace*{2em}\mbox{ at } r= - 1$$
and
$$T= 1 \hspace*{2em}\mbox{ at } r=  1.$$

\subsection{Numerical approach}

A pseudo-spectral collocation-Chebyshev and Fourier method is implemented. In the
meridional plane $(r,z)$, each dependent variable is expanded in the approximation
space $P_{NM}$, composed of Chebyshev polynomials of degrees less than or equal to
$N$ and $M$ respectively in $r$- and $z$- directions, while Fourier series are
introduced in the azimuthal direction.

Thus, we have for each dependent variable $f$ ($f=V_r, V_\phi, V_z, T$):
\begin{equation}
 f_{NMK}(r,\phi,z,t) = \sum_{n=0}^{N}\,\sum_{m=0}^{M}\,\sum_{k=-K/2}^{K/2-1}\, \;{\hat f}_{nmk}(t) T_n(r)
 T_m(z) exp(ik \phi)
\end{equation}
where $T_n$ and $T_m$ are Chebyshev polynomials of degrees $n$ and $m$.

This approximation is applied at collocation points, where the differential
equations are assumed to be satisfied exactly (\cite{gott77}, \cite{canu87}). We
have considered the Chebyshev-Gauss-Lobatto points,
\begin{eqnarray*}
r_i & = & \cos(\frac {i \pi}{N}) \mbox{ for } i \in [0,N],\\
z_j & = & \cos(\frac {j \pi}{M}) \mbox{ for } j \in [0,M],
\end{eqnarray*}
and a uniform distribution in the azimuthal direction:
$$
\phi_k = \frac{2k \pi}{K} \mbox{ for } k \in [0,K[.
$$

The time integration used is second order accurate and is based on a combination of
Adams-Bashforth and Backward Differentiation Formula schemes, chosen for its good
stability properties (\cite{van86}). The resulting AB/BDF scheme is semi-implicit,
and for the transport equation of the velocity components in (\ref{NSE1}) can be
written as:

\begin{eqnarray} \label {EDV}
\frac{3{f}^{l+1}-4{f}^l+{f}^{l-1}}{2 \delta t} +
 2\mathcal{N}({f}^l) - \mathcal{N}({f}^{l-1}) \nonumber \\
 = -\nabla \Pi^{l+1} + C \nabla^2 {f}^{l+1} + F_i^{l+1}
\end{eqnarray}

where $\mathcal{N}(f)$ represents nonlinear terms with the Coriolis term, $F_i$ corresponds to
the component of the forcing term $\mathbf{F}$, $\delta t$ is the time step and the
superscript $l$ refers to time level. The cross terms in the diffusion part in the $(r, \phi)$
plane resulting from the use of cylindrical coordinates in  (\ref{NSE1}) are treated within
$\mathcal{N}(f)$, in order to maintain an overall second order time accuracy. An equivalent
discretisation applies for the transport equation of the temperature (\ref{NSE3}). For the
initial step, we have taken ${f}^{-1} = {f}^{0}$. At each time step, the problem then reduces
to the solution of Helmholtz and Poisson equations.

An efficient projection scheme is introduced to solve the coupling between the velocity and
the pressure in (\ref{NSE1}). This algorithm ensures a divergence-free velocity field at each
time step, maintains the order of accuracy of the time scheme for each dependent variable and
does not require the use of staggered grids (\cite{hugu98}, \cite{raspo02}). A complete
diagonalisation of operators yields simple matrix products for the solution of successive
Helmholtz and Poisson equations at each time step (\cite{hald84}). The computations of
eigenvalues, eigenvectors and inversion of corresponding matrices are done once during a
preprocessing step.

\subsection{Computational details}
The numerical code has been previously validated by \cite{RaLeBo98} for a liquid-filled cavity
($Pr=13.07$) by comparison of solutions with the detailed results reported by \cite{Hetal85}
from a combined laboratory and numerical study presented in \cite{Hignett85}. Very close
agreement has been obtained between our results and measurements for a regular wave flow.
Particular attention has been paid to the grid effect on the solution, which has served as
basis for the present study.  A grid independency of the three-dimensional solution is assumed
when the levels of non-harmonics of the dominant wave amplitude approach the machine precision
($10^{-9}$).

For $Ta\le 3\times 10^5$, a mesh of $N \times M \times K = 64 \times 96 \times 80$ was used in
the radial, axial, and azimuthal directions, respectively, with a dimensionless time step
$\delta t = 0.1125$. For $Ta > 3\times 10^5$, a finer resolution was used, $N \times M \times
K = 96 \times 96 \times 128$, and various time steps within the numerical stability limit: a
dimensionless time step, $0.003225\leq \delta t \leq 0.05$.

For the transition from the upper symmetric regime to the regular waves, the initial
conditions corresponded to the steady axisymmetric solution at each azimuthal node to which a
random perturbation was added to the temperature field in azimuth. Subsequently, the strategy
consisted of increasing or decreasing progressively the rotation rate, without adding any
further perturbations, for following successive three-dimensional solutions.

\section{Data analysis}
The data analysis techniques can be broadly classified into two-dimensional sections for
visualisation of the fields, spectral analysis of one-dimensional sections for mode amplitudes
and frequencies, and phase space reconstruction from dynamical systems theory to provide
information on the qualitative class of solution observed.

\subsection{Two-dimensional fields and azimuthal sections}
To visualise the flow, both horizontal and vertical cross-sections of the velocity,
temperature, and pressure fields were used.  The horizontal sections shown here are at
mid-height, and the vertical cross-sections are in the $r,z$-plane at zero azimuthal angle.

For the further parametric analysis, the sequence of instantaneous three-dimensional fields
was reduced to time series of mode amplitudes in a one-dimensional section.  These could then
be used for spectral analyses and phase space reconstructions using techniques from dynamical
systems theory.  The sections considered here are azimuthal sections (rings) at mid-height
and, unless specified otherwise, at mid-radius.  These sections then gave the temporal
evolution of the set of azimuthal Fourier modes.  To obtain a measure of the radial structure,
other sections were taken also at mid-height but at a quarter of the gap width from the inner
wall and a three quarters, respectively.  The normalised space variables were calculated as
the sum of contribution from the Chebyshev polynomials, evaluated at $(r,z)=(0,0)$ for each
azimuthal Fourier mode for the mid-height, mid-radius section, and at $(r,z)=(-0.5,0)$ and
$(0.5,0)$ at the one and three quarter sections, respectively.

\subsection{Spectral analysis}
From the time series of the variables in the azimuthal sections one can obtain an
instantaneous or time-averaged spatial spectrum of the azimuthal wave number to calculate both
the average relative amplitudes of each azimuthal wave mode and to show the temporal evolution
of each mode.  Each Fourier mode contains information on both the spatial phase and the
amplitude of a particular wave solution.  If amplitude variations are much smaller than the
mean amplitude of a wave, the signal in a Fourier mode is dominated by the drift of that mode
in the azimuthal direction. By taking the square-root of the sum of squares of each pair of
sine and cosine modes, only the wave amplitude is retained, making it possible to investigate
small amplitude fluctuations.  Both, Fourier mode time series and wave amplitude time series,
were used to calculate power spectra.

\subsection{Phase space reconstruction and analysis}\label{phaseanalysis}
Two types of phase portraits and corresponding Poincar\'e sections were generated, one using
the Fourier modes as the basis functions and the other using the wave amplitudes.  Both types
of phase portraits and Poincar\'e sections produced qualitatively equivalent results for each
regime, but the analysis using the wave amplitudes effectively halved the dimension of the
phase space and proved effective in analysing flow of a higher attractor dimension.

The largest Lyapunov exponent $\lambda_1$ in a number of cases was estimated using the method
of \cite{WSSV85} but using the univariate Singular Systems Analysis (SSA) by \cite{BrKi86}
method to obtain a cleaner embedding of the attractor.  The SSA analysis is essentially a
singular value decomposition of the covariance matrix obtained from the Takens delay matrix,
and it concentrates the variance in the signal in a few leading singular vectors.
\cite{RBJS92}, using experimental measurements, had found this to be an effective compromise
between optimal attractor reconstruction and computational expense, and has proven a highly
effective means of detecting the onset of chaotic behaviour as $\lambda_1$ becomes
significantly positive. The method estimates the largest Lyapunov exponent from long time
series, by tracking the divergence of nearby trajectory segments on the reconstructed
attractor for a specified time period, the evolution time $\tau_E$, and averaging the
logarithmic divergence rate over a large ensemble of segments covering the entire
reconstructed attractor. The robustness of the resulting Lyapunov exponent estimate was tested
by repeating the calculation using a range of assumed embedding dimension, the tolerance of
initial perturbations and/or evolution time $\tau_E$.

\section{Results} \label{Results}
\subsection{The regimes observed} \label{regimes}
\begin{figure}
\includegraphics[width= \textwidth]{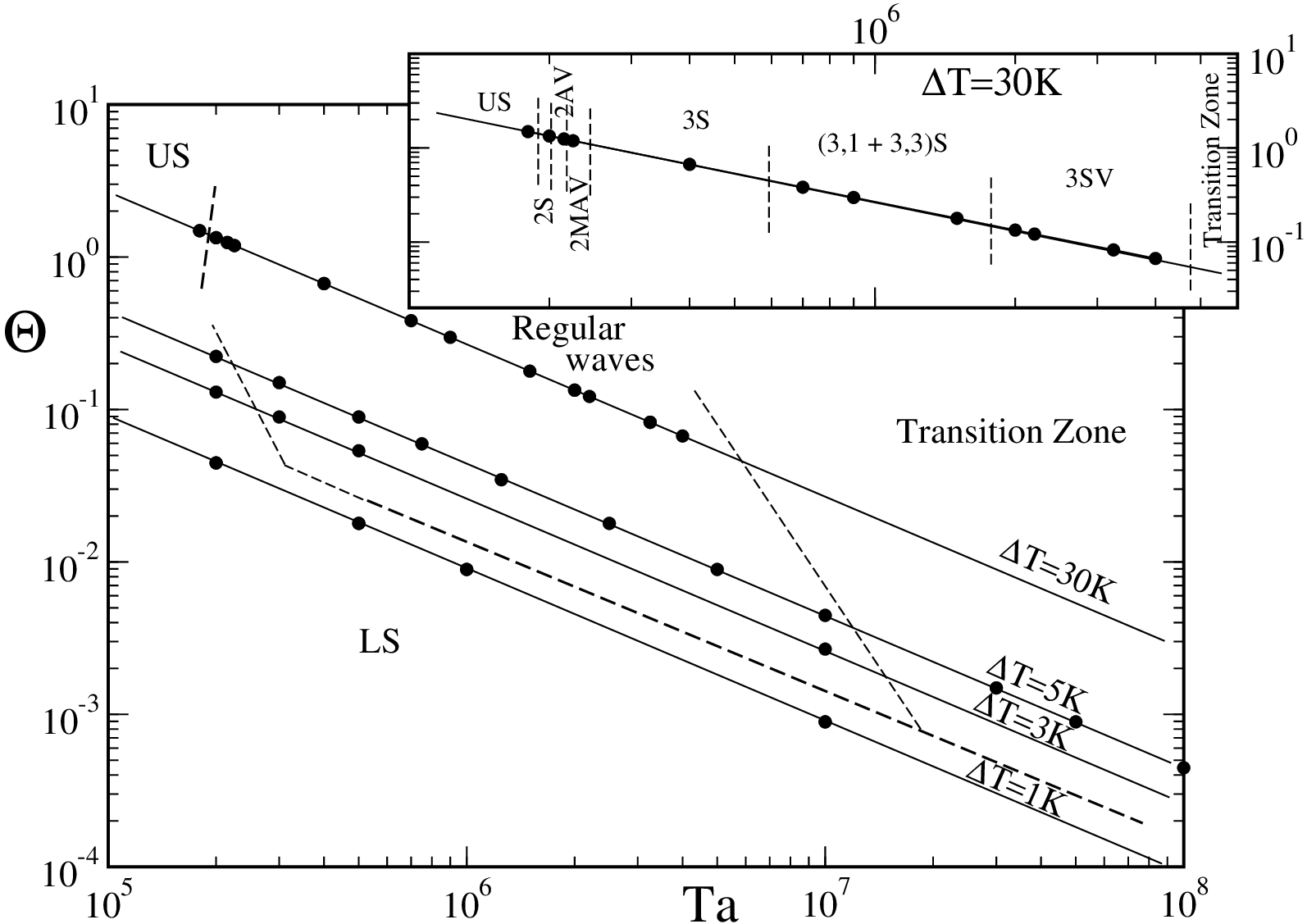}
\caption{\label{Regime}Summary of the regimes for air with Prandtl number of 0.7 as obtained
by the numerical model.  The solid lines show the temperature differences investigated in this
paper and previously by \cite{MaRa02} and \cite{MaRa03}. The dots on the lines for $\Delta T=
1, 3, 5\rm K$ correspond to cases discussed in the earlier papers while the dots on the line
for $\Delta T= 30\rm K$ show the cases discussed here.  The dashed line indicates the
transition from the symmetric regimes to regular waves, and the dotted line indicates the
transition from regular waves to irregular flow.   The line for $\Delta T= 30\rm K$ is shown
again enlarged in the insert to identify the different regimes observed in this study.}
\end{figure}
One of our objectives was to delineate a detailed regime diagram for air which was as complete
as possible, similar to those already available in literature for various liquids. In Figure
\ref{Regime} we present a coarse-grained regime diagram obtained from direct simulation of air
inside the same geometry as used by \cite{FH65} in their experiments with liquids. This regime
diagram uses and extends initial investigation primarily concerned with the delimitation of
the axisymmetric regimes by \cite{MaRa02} and \cite{MaRa03}.

The regime where all perturbations decay to an axisymmetric flow is traditionally split into a
`{\it Lower Symmetric}' (LS) regime, where the flow is strongly affected by viscous diffusion,
and an `{\it Upper Symmetric}' (US) regime, where the flow is stabilised by stratification due
to the background rotation.  Despite the different labels and processes involved, there exists
no clear phase transition separating them. At $\Delta T = 1K$, only the LS axisymmetric
solution was found for all values of rotation rates considered up to a Taylor number of
$Ta=10^8$. At higher temperature differences $3K \le \Delta T \le 30K$, we have delineated the
transition between the symmetric regimes and regular waves. In the US regime, characterised by
high values of the thermal Rossby number (and small values of the Taylor number), the heat
transfer is still dominated by conduction. This transition occurs following the criterion
reported by \cite{hide58} from his experimental observations: $\Theta \le \Theta_{crit} \equiv
1.58 \pm 0.05$\@. Moreover, \cite{hide58} mentioned that this anvil-like curve separating the
upper symmetric regimes and the regular waves does not significantly vary with the fluid
properties, say the Prandtl number $Pr$, unlike the lower symmetric transition boundary,
e.g.~\cite{FP76}.  The computations were not extended beyond $\Delta T=30\rm K$, partly for
practical reasons and partly because the validity of the Boussinesq approximation may become
debatable when the volume expansion coefficient, $\Delta T/T_0$, becomes much larger than
$0.1$\@. As a result, we have only confirmed the knee of the anvil in Figure~\ref{Regime} but
not the upper, almost horizontal section of the traditional anvil.

At $\Delta T = 5K$, a transition to irregular waves has been obtained through the route:\\
\emph{Upper symmetric flow $\rightarrow$ steady waves $\rightarrow$
 quasi-periodic waves $\rightarrow$ irregular flow}
(\cite{MaRa02}). By convention, the term 'steady wave' is used to identify a flow
characterised by a wave of constant amplitude and drifting at a constant speed in the
azimuthal direction.  Substantial hysteresis was found as a result of the coexistence of
different wave flow solutions under the same physical conditions, consistent with rotating
experiments in liquids and many other systems of rotating flows in confined configurations.
This phenomenon is also known as intransitivity.  In the present case, each solution resulted
from adding to the axisymmetric solution a small perturbation of a selected wavenumber to the
temperature field.  However, the maximum wavenumber obtained remained within the empirical
limits found by \cite{hide58}: $m_{max} \le 0.67 \pi R_c \approx 8$ for the present geometry.
During this preliminary exploration, we did not identify clearly any vacillation regimes for
$\Delta T \le 5K$.

Some vacillation was observed at a higher temperature difference of $\Delta T = 30K$ in a
first exploration of this temperature contrast by \cite{MaRa03}.   The work reported here
continued these observations in a detailed investigation and analysis of the flow regimes
obtained for this value of $\Delta T$. The specific cases discussed in detail are summarised
in Table~\ref{tablesummary}.

\begin{table}
\centering
\begin{tabular}{c c c c c l l l l}
 $Ta (\times 10^{5})$       &   $\Theta$   &  $\Omega$ (rad/s)
                     & regime       & $m$  & $\omega_d$ ($2\Omega$) &
                     $\omega_{v,m}$ ($2\Omega$) & $A_{m,1}, A_{m,3}$ & $\eta$  \\ \hline
 1.775 & 1.5087  &  3.617  & $US$         & 0 &          &         &   & \\
 1.800 & 1.4877  &  3.642  & $US$         & 0 &          &         &   &  \\
 \hline
 1.800 & 1.4877  &  3.642  & $2S$         & 2 &          &         & 0.16666 & \\
 1.825 & 1.4673  &  3.667  & $2S$         & 2 & 0.004439 &         & 0.18335 &     \\
 1.850 & 1.4475  &  3.693  & $2S$         & 2 & 0.003104 &         & 0.19533 &     \\
 2.000 & 1.3389  &  3.839  & $2S$         & 2 & 0.004113 &         & 0.24676 &     \\
 2.050 & 1.3063  &  3.887  & $2S$         & 2 & 0.006102 &         & 0.26097 &     \\
 2.075 & 1.2905  &  3.911  & $2AV$        & 2 & 0.006647 & 0.05806 & 0.26137 & 0.0161  \\
 2.100 & 1.2752  &  3.934  & $2AV$        & 2 & 0.00704  & 0.05770 & 0.2617  & 0.0343  \\
 2.150 & 1.2455  &  3.981  & $2AV$        & 2 & 0.00740  & 0.05676 & 0.2630  & 0.0692  \\
 2.175 & 1.2312  &  4.004  & $2AV$        & 2 & 0.00755  & 0.05630 & 0.2639  & 0.0845  \\
 2.200 & 1.2172  &  4.027  & $2MAV$       & 2 & 0.00783  & 0.04925,& 0.2684  & 0.1448  \\
       &         &         &              &   &          & 0.002344 &        &         \\
 2.250 & 1.1902  &  4.072  & $2MAV$       & 2 & 0.00829  & 0.04398,& 0.2782  & 0.2589  \\
       &         &         &              &   &          & 0.0037  &         &         \\
 2.2925& 1.1681  & 4.110   & $2MAV$       & 2 & 0.00893  & 0.04206,& 0.2895  & 0.2993  \\
       &         &         &              &   &          & 0.0070  &         &         \\
 \hline
 2.000 & 1.3389  &  3.839  & $3S$         & 3 &          &         & 0.2700  &      \\
 2.150 & 1.2455  &  3.981  & $3S$         & 3 &          &         & 0.2991  &      \\
 2.250 & 1.1902  &  4.072  & $3S$         & 3 &          &         & 0.3180  &      \\
 \hline
 2.300  & 1.1643 & 4.117   & $3S$         & 3 & 0.00905  &         & 0.3269  &      \\
 3.000  & 0.8926 & 4.702   & $3S$         & 3 & 0.01551  &         & 0.3612  &      \\
 4.000  & 0.6695 & 5.430   & $3S$         & 3 & 0.01964  &         & 0.4200  &      \\
 5.000  & 0.5356 & 6.070   & $(3,1+3,3)S$ & 3 & 0.02094  &         & 0.4446, &      \\
       &         &         &              &   &          &         & 0.0721  &      \\
 7.000  & 0.3826 & 7.183   & $(3,1+3,3)S$ & 3 & 0.02244  &         & 0.4524, &      \\
       &         &         &              &   &          &         & 0.1500  &      \\
 9.000  & 0.2975 & 8.144   & $(3,1+3,3)S$ & 3 & 0.02513  & 0.1839  & 0.4469, &      \\
       &         &         &              &   &          &         & 0.1918  &      \\
 15.000 & 0.1785 & 10.514  & $(3,1+3,3)S$ & 3 & 0.02732  & 0.1963  & 0.4152, &      \\
       &         &         &              &   &          &         & 0.2280  &      \\
 20.000 & 0.1339 & 12.141  & $3SV$        & 3 & 0.02856  & 0.2183, & 0.3937, & 0.004131 \\
       &         &         &              &   &          & 0.6367, & 0.2297  &      \\
       &         &         &              &   &          & 0.0309  &         &      \\
 22.000 & 0.1217 & 12.734  & $3SV$        & 3 & 0.02857  & 0.1916, & 0.3863, & 0.009798 \\
       &         &         &              &   &          & 0.6465, & 0.2265  &      \\
       &         &         &              &   &          & 0.0333  &         &      \\
 32.500 & 0.0824 &  15.477 & $3SV$        & 3 & 0.02946  & 0.1891, & 0.3598, & 0.078200 \\
       &         &         &              &   &          & 0.567,  & 0.2185  &      \\
       &         &         &              &   &          & 0.09    &         &      \\
\end{tabular}
\caption{Summary of computed results for $\Delta T=30K$, with $m$ the dominant wave number.}
\label{tablesummary}
\end{table}

\subsection{First instability and transition to steady waves}\label{m2}
In this section, the initial transition from the upper symmetric regime (US) to a
steady wave with azimuthal wave number $m=2$ (2S), on increase of the Taylor number,
will be presented.  This will be followed by discussion of the subsequent
transitions to an Amplitude Vacillation (2AV) and the more complex so-called
`Modulated Amplitude Vacillation' (2MAV; cf \cite{RBJS92} and \cite{FR97}).

\begin{figure}
\centering
\includegraphics[width= 0.8\textwidth]{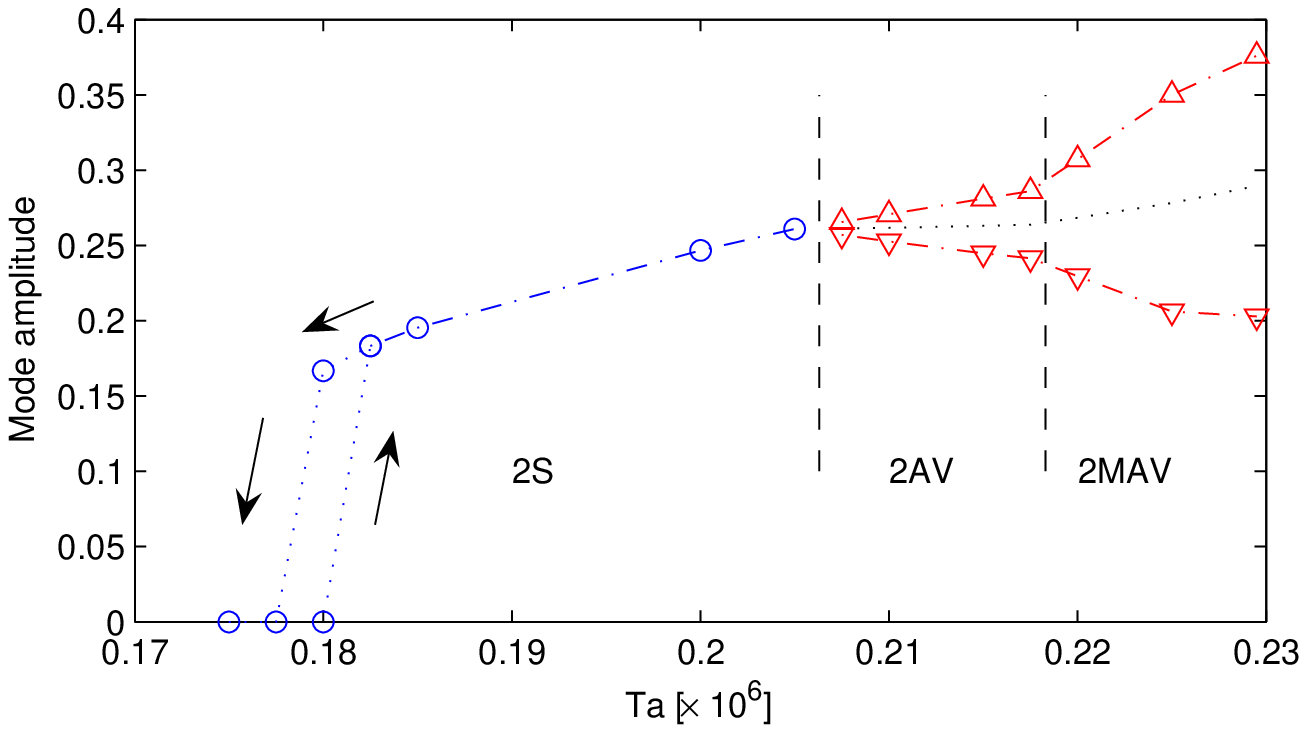}
\caption{\label{Amplitude} The amplitude of the dominant azimuthal mode, $m=2$, at mid-radius.
The dash-dotted lines indicate the global maximum and minimum of the amplitude. The
dash-dotted lines in the 2MAV regime denote the largest amplitude minimum and the smallest
amplitude maximum of all the extreme values in each vacillation cycle, while the dotted lines
in 2MAV show the average of the amplitude minimum and maximum for all vacillation cycles,
respectively.  The dotted line in the 2AV regime shows the averaged amplitude.}
\end{figure}
\begin{figure}
\includegraphics[width=\textwidth]{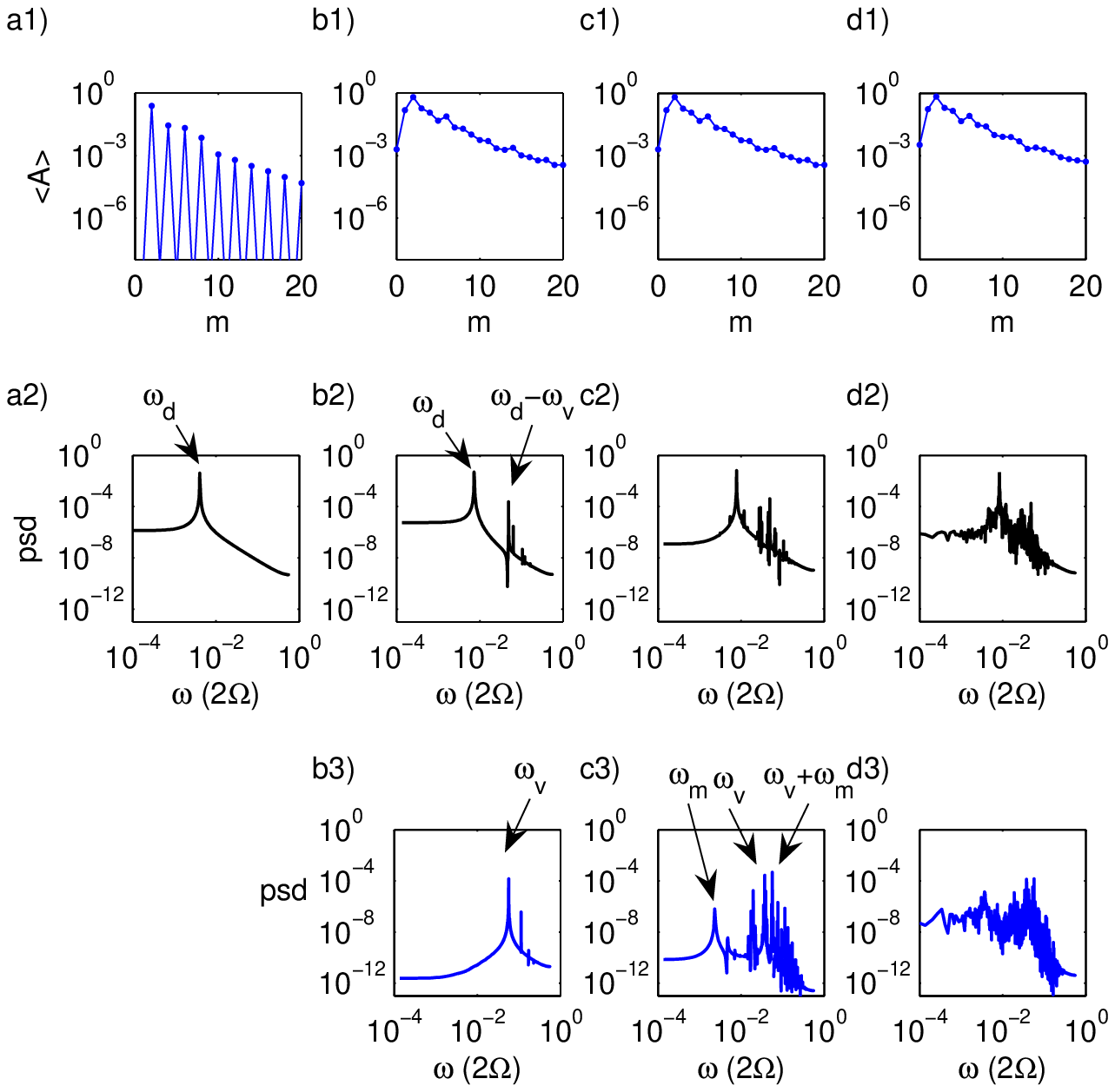}
\caption{\label{Spectra2} Spectra of representative cases of the $m=2$ flows from time series
of the amplitudes of the azimuthal modes at mid-radius and mid-height. Column a) steady wave
2S, $Ta= 200,000$, column b) Amplitude vacillation 2AV, $Ta= 215,000$, column c)
quasi-periodic modulated amplitude vacillation, 2MAV, $Ta=220,000$, and column d) irregular
2MAV, $Ta=225,000$.  Top row 1): spatial spectra of the time-averaged amplitude of the
azimuthal wave mode, $m$. Middle row 2) temporal spectra of the time series of the cosine
component of mode $m=2$.  Bottom row 3): temporal spectra of the amplitude of $m=2$.  The main
frequencies are indicated by their annotation, $\omega_d$ for the drift frequency, $\omega_v$
for the vacillation frequency, and $\omega_m$ for the modulation frequency.}
\end{figure}

The range of regimes dominated by the azimuthal wave mode $m=2$ is summarised in
Figure~\ref{Amplitude}, which shows the equilibrated or time-averaged amplitude of that mode.
The axisymmetric baroclinic circulation was found to become unstable when the Taylor number
was increased from $Ta= 180,000$ to $182,500$, following which the solution converged to a
steady wave mode $m=2$ and its harmonics, as a time-averaged spatial spectrum at mid-height
and mid-radius demonstrates for a qualitatively similar case in Figure \ref{Spectra2}(a1). The
corresponding temporal spectrum in Fig.~\ref{Spectra2}(a2) of the cosine component of mode
$m=2$ shows that the flow consists of a steady wave of constant amplitude travelling at a
fixed angular velocity around the tank.  The amplitude of the first wave solution and the step
to it from the axisymmetric solution are indicated in the portion of Figure~\ref{Amplitude}
indicated by 'US' and '2S'.  The regions identified by '2AV' and '2MAV' correspond to
time-dependent regimes discussed below in section~\ref{sAV}, and the symbols represent the
minima and maxima of the wave amplitude observed over time.

Decreasing the Taylor number from this first wave solution at $Ta=182,500$ back to
$Ta=180,000$ did not result in a return to the axisymmetric solution but to another steady
wave solution of slightly lower amplitude and higher angular velocity, as indicated in Figure
\ref{Amplitude}.  This hysteresis extended over a small range of Taylor number only, so that
reducing the Taylor number still further to $Ta=177,000$, for example, resulted in a slow
decay of the wave back to the axisymmetric solution.

\subsubsection{Basic flow structures}
\begin{figure}
(a) \hspace{0.48\textwidth}(b)\\
\includegraphics[width=0.47\textwidth, height=0.47\textwidth]{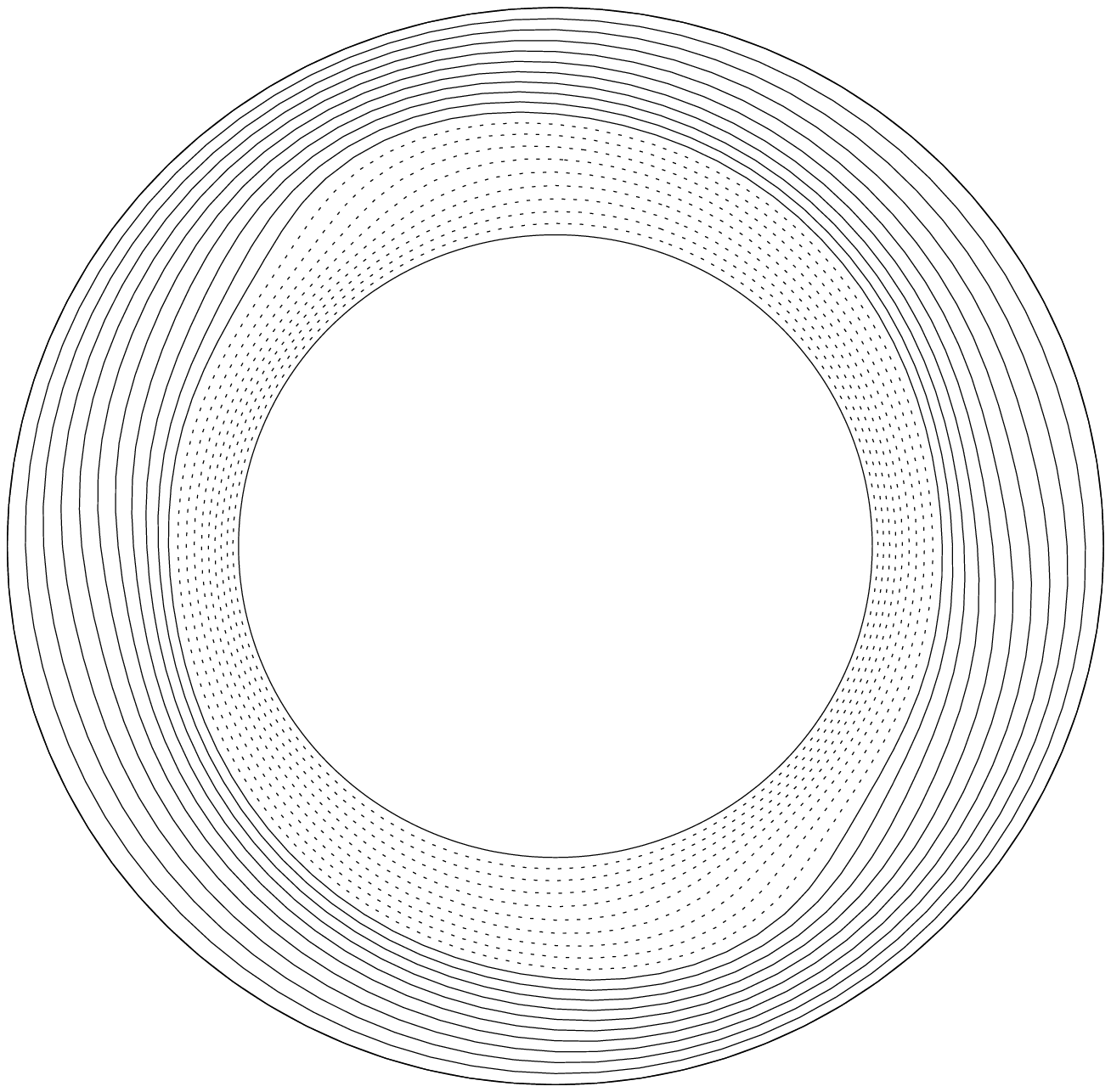}
\hfill
\includegraphics[width=0.47\textwidth, height=0.47\textwidth]{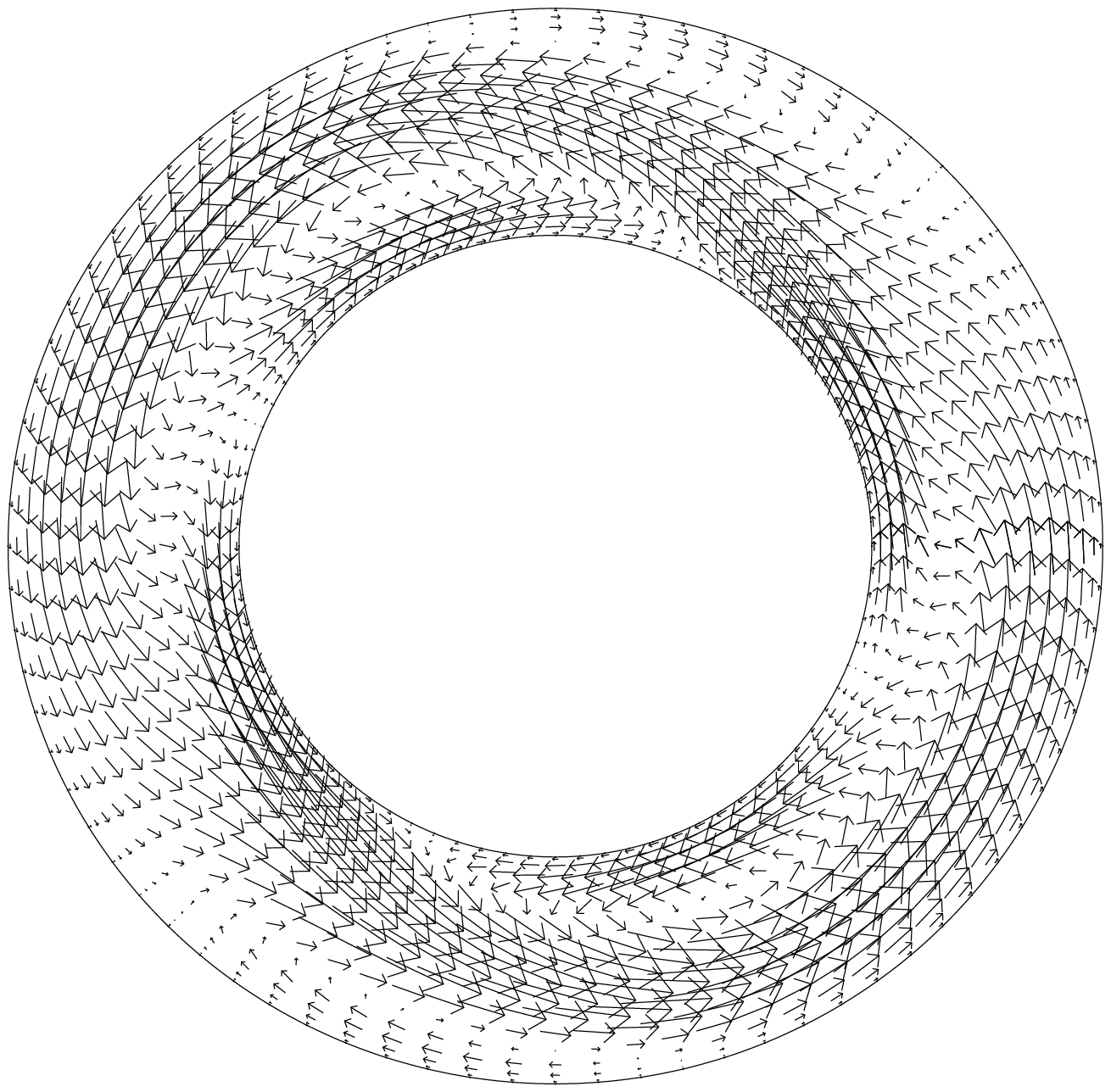}
\caption{\label{Field2} Horizontal cross-sections at mid-height of a 2S flow at $Ta=200,000$.
(a) Temperature contours and (b) the horizontal velocity field.}
\end{figure}
\begin{figure}
\includegraphics[width=\textwidth]{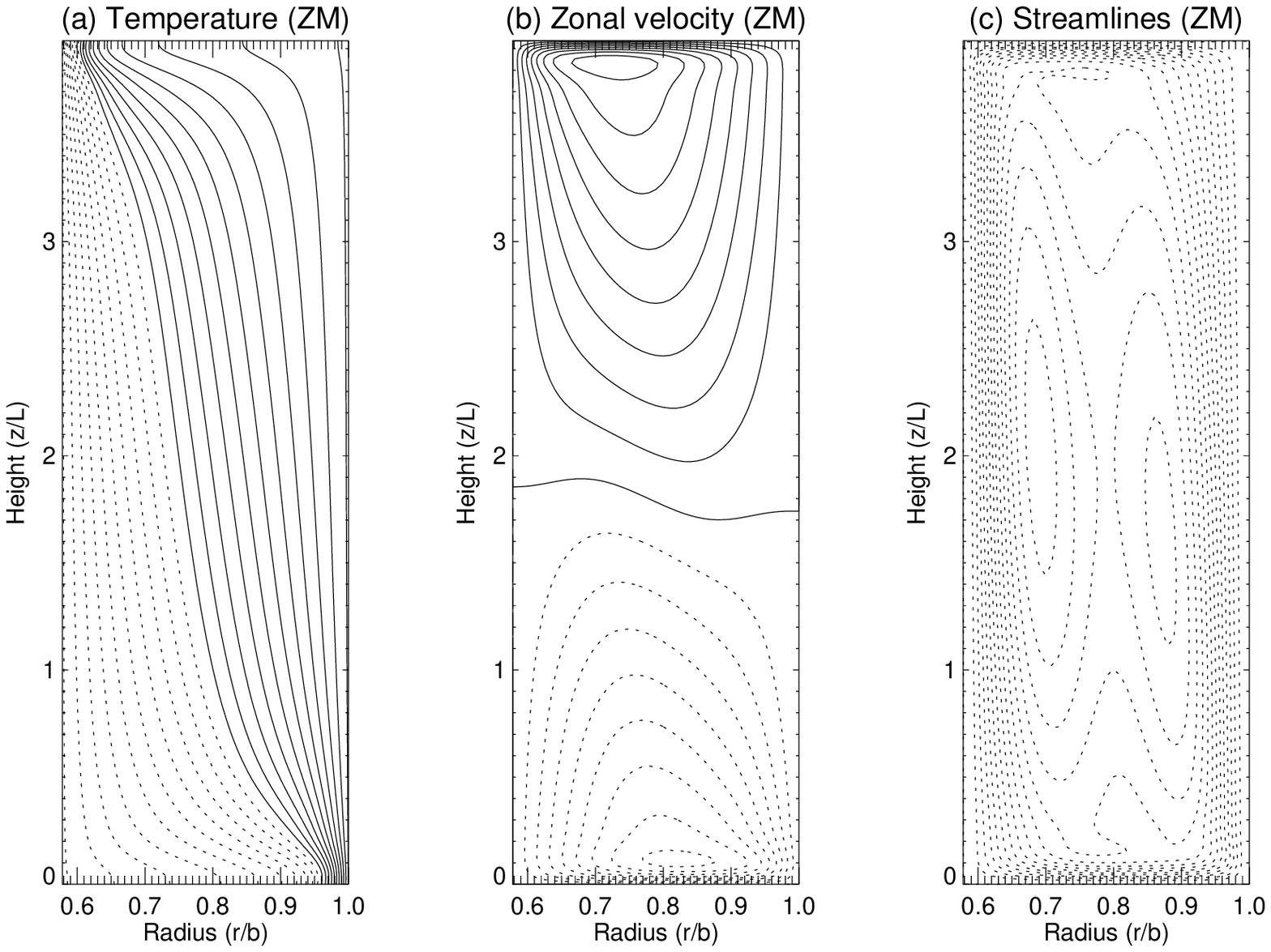}
\caption{\label{ZMfields1} Zonal mean fields at $Ta = 200,000$ using air as the working fluid.
(a) Temperature (contour interval = 0.1, normalised to an imposed temperature difference of
$\pm 1.0$), (b) zonal velocity $V_\phi$ (contour interval = 0.25) and (c) meridional stream
function (contour interval = 0.01). }
\end{figure}
\begin{figure}
\includegraphics[width=\textwidth]{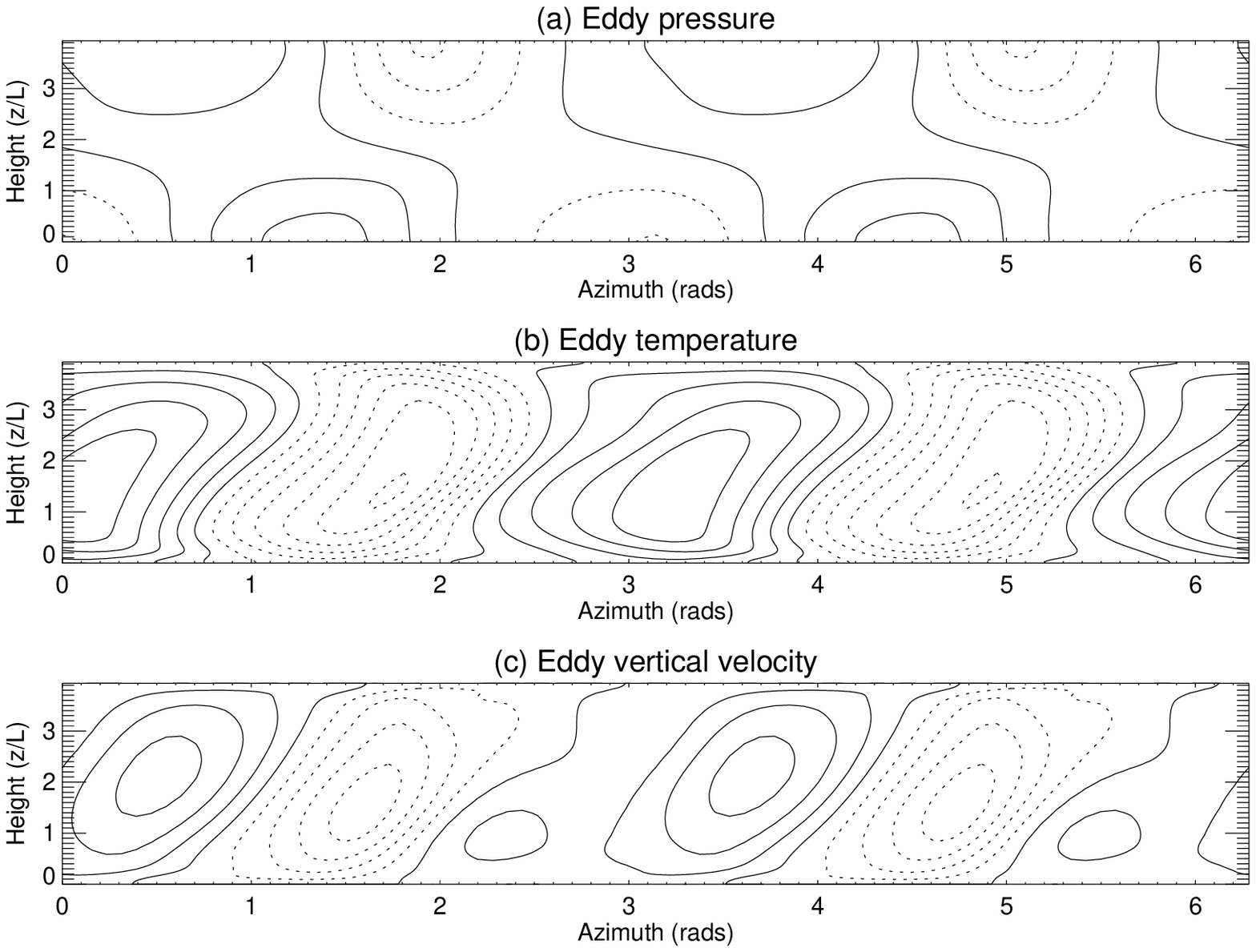}
\caption{\label{TZfields1} Azimuth-height maps of eddy fields (with azimuthal flow removed) in
a typical simulation of a 2S flow at $Ta = 200,000$ using air as the working fluid. (a)
Pressure (contour interval = 0.05 dimensionless units), (b) Temperature $T$ (contour interval
= 0.05, normalised to an imposed temperature difference of $\pm 1.0$) and (c) Vertical
velocity (contour interval = 0.1 dimensionless units). }
\end{figure}

An example of a typical $m=2$ flow is presented in Figures \ref{Field2} to \ref{TZfields1} of
a steady wave, 2S, at a Taylor number of $Ta=200,000$. Figure \ref{Field2} shows temperature
contours and the horizontal velocity field at mid-level. These show the breaking of the
azimuthal symmetry into a discrete symmetry of a mode with azimuthal wave number $m=2$. The
velocity field can be described as a meandering jet stream, with alternating cyclones and
anticyclones on either side of the jet stream. The presence of such a jet at mid-level is
indicative of a significant breaking of the antisymmetry of the azimuthal flow about the
mid-plane.

The basic, azimuthally-averaged flow associated with the $m = 2$ waves is illustrated in
Fig.~\ref{ZMfields1}. This shows some differences from the typical azimuthal mean state found
in experiments using liquids of Prandtl number $\gg 1$. In particular, the temperature field
shows relatively weakly-developed sidewall thermal boundary layers and isotherms which are
steeply sloped and nearly vertical in places.  On the other hand, the velocity fields and
azimuthal mean stream lines such as in Fig.~\ref{ZMfields1}(b) and (c) indicate that Ekman
layers are well developed on the rigid, horizontal boundaries while somewhat thicker, viscous
Stewartson layers are present at both sidewalls.

The baroclinic character of the waves is illustrated in Fig.~\ref{TZfields1}, which displays
azimuth-height sections at mid-radius but with the azimuthal mean flow removed. The pressure
field in Fig.~\ref{TZfields1}(a) exhibits the strong `westward' phase tilt with height
characteristic of baroclinically unstable waves, e.g.~\cite{HM75}. The corresponding
temperature field, $T'$ in Fig. \ref{TZfields1}(b), shows a strong $m = 2$ disturbance with a
noticeable \emph{eastward} phase tilt over most of the height range, in qualitative
resemblance to the classical, linearly unstable Eady wave, e.g.~\cite{HM75}. The vertical
velocity field in Fig. \ref{TZfields1}(c) is closely correlated with that of $T'$, indicating
that the fully-developed wave flow has optimised the baroclinic release of potential energy in
the buoyancy field by ensuring that the upward eddy heat flux $\overline{w'T'}$ is as large as
possible, where the $w'$ and $T'$ are the deviations from the temporal mean and the overbar is
the time averaging of their product. The azimuthally asymmetric nature of this $m = $2S flow
is again clearly apparent, but in other respects this initial instability evidently develops
as a fairly `classical' form of baroclinic wave with clear similarities to those found for $Pr
\gg 1$ (allowing for the differing balance of diffusive effects at $Pr \sim 1$).

\subsubsection{Transitions to Amplitude Vacillations}\label{sAV}
The steady wave solution remained qualitatively the same for increased Taylor number until
after $Ta=205,000$, where the wave amplitude increased gradually but the angular velocity
first decreased and then increased again.  At $Ta= 207,500$, a noticeable periodic oscillation
of the amplitude was observed, exhibiting the characteristic signature of a simple Amplitude
Vacillation, 2AV in this case.  Figure \ref{Spectra2}~(b1) shows that the time-averaged
spatial wave spectrum for the AV is fundamentally different from the steady waves, with
substantial energy in wave modes other than the dominant wavenumber and its harmonics.  The
temporal spectra of mode $m = 2$ in Figure \ref{Spectra2}~(b2 and b3) now show a second free
frequency in the spectrum of the cosine component (b2) and an associated peak in the spectrum
of the mode amplitude (b3). These two peaks are at different frequencies because the peak in
the cosine spectrum is the difference between the vacillation and drift frequencies.

The magnitude of that vacillation is indicated by the extreme values for the wave amplitude in
Figure~\ref{Amplitude} (a), where the envelope of the vacillation is shown by the maximum
amplitude (the upward pointing triangles) and the minimum amplitude (the downward pointing
triangles).  It is quantified by the vacillation index,
$\eta=(A_{max}-A_{min})/(A_{max}+A_{min})$, where $A_{max}$ and $A_{min}$ are the amplitude
maximum and minimum within a vacillation cycle, respectively. This vacillation index increased
first gradually then more steeply. This sharp increase in the vacillation index occurred
together with the onset of a slow modulation of the vacillation, 2MAV, adding a third free
frequency, as seen in Figure \ref{Spectra2}~(c3).  While that spectrum is still fairly clear
with sharp peaks, the spectra at $Ta= 225,000$ and $229,250$ are much broader.  The spatial
spectra of the MAV flows as shown in Figures~\ref{Spectra2}~(c1 and d1), however, are
indistinguishable from that of the 2AV flow.

An important and novel point to note here is that the progression from a steady wave
to an AV flow on increase of the rotation rate of Taylor number has not usually been
seen in most laboratory experiments using fluids with a Prandtl number of seven and
higher (e.g. \cite{Hetal85,RBJS92,KoWh81}). In their cases, the transition to AV was
typically found to occur on \emph{decrease} of the rotation rate (or $Ta$), where it
has been observed that the transition seems to occur earlier the higher the Prandtl
number (\cite{J81})\footnote{\cite{Hignett85} has discussed that some of the
classification by \cite{J81} was incorrect.  However, a comprehensive synopsis of
the rotating annulus literature confirms this phenomenon.}. An exception at high
Prandtl number seems to be found close to the lower symmetric transition boundary
(e.g.~\cite{LN04}), where the onset of AV appears via a straightforward
supercritical secondary Hopf bifurcation with respect to $Ta$ or $\Theta$, much as
found here for air. The present results thus indicate important roles for $Pr$ and
$Ta$ in determining the nature of the bifurcation to quasi-periodic AV either as
super- or sub-critical with respect to $Ta$ or $\Theta$.

\subsubsection{Phase portrait analysis}
\begin{figure}
\includegraphics[width=\textwidth]{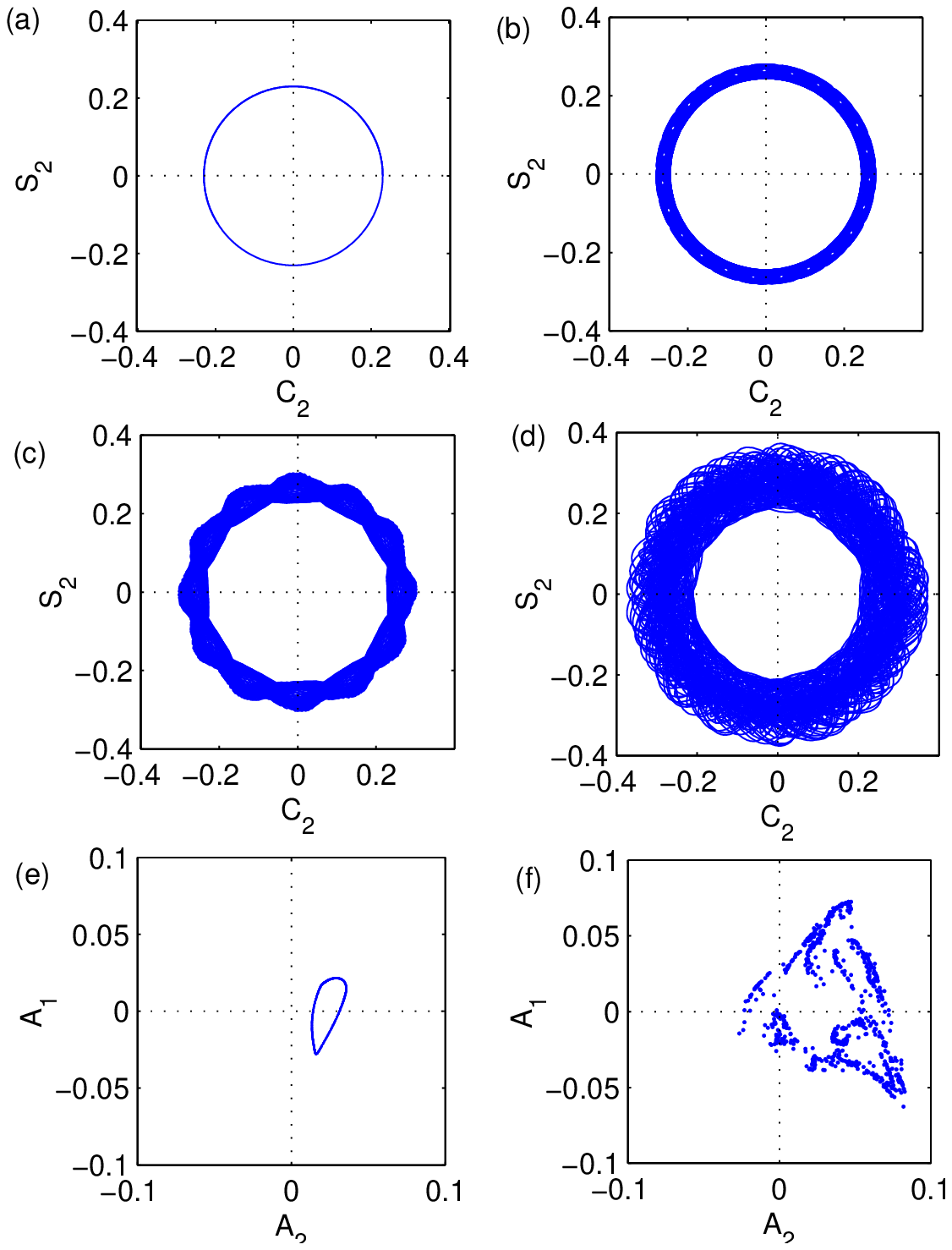}
\caption{\label{Delay2}Phase space representations of: (a) a steady 2S (Ta= 200,000), (b) a
2AV (Ta=215,000), (c) a quasi-periodic 2MAV (Ta= 220,000), and (d) a chaotic 2MAV
(Ta=229,250). The phase space diagrams show the time series of the sine component against that
of the cosine component of mode $m=2$.  Panels (e) and (f) show Poincar\'e sections of the
amplitudes of $m=1$ against $m=2$ where $m=4$ is zero, after the mean amplitudes had been
subtracted. }
\end{figure}
The trajectory in a reconstructed phase space or configuration space provides a way to
demonstrate the different qualitative behaviours of the different flow regimes. Two
representations were used here and are shown in Figure~\ref{Delay2}.   One representation uses
the azimuthal sine and cosine modes of a Fourier analysis of the fields at mid-height and
mid-radius, the other uses the amplitude of the modes thereby removing the phase information
of the wave. The trajectories, or orbits, in the projection onto the cosine and sine
components of mode $m=2$ at mid-radius are shown in Figure~\ref{Delay2} (a to d) for the
different $m=2$ flows, 2S (a), 2AV (b), and two 2MAV cases (c and d). The simple circle in (a)
represents the angular progression in azimuth of a wave with constant amplitude. The 2AV case
in (b) shows a thickening of the circle which a Poincar\'e section revealed to be a torus. The
large circle dominating in the phase plot in (b) still represents the drift of the wave while
the small circle of the torus cross-section describes the periodic oscillation of the wave
amplitude in the Amplitude Vacillation.  Figure~\ref{Delay2} (c) still shows some structure
but with more detail which cannot be attributed to a simple oscillation on a torus, and (d)
only retains the obvious structure of the wave mode traveling in azimuth.

If the spatial phase information is separated from the amplitude information, the amplitudes
of different modes can also be used to construct an attractor of reduced dimension in a
pseudo-phase space.  This in effect removes the drift and its frequency from the display, the
circle in Figure~\ref{Delay2} (a) turns into a fixed point, the torus in (b) into a limit
cycle, and a 3-torus into 2-torus.  To analyse the 2MAV flow, a Poincar\'e section of the
amplitude trajectory is required.  A section to the 2MAV case at $Ta=220,000$ in
Figure~\ref{Delay2}(e) reveals that the orbit in the amplitude space is a 2-torus, and
therefore that the 2MAV shown in (c) is a 3-torus. This implies that this particular 2MAV flow
is a quasi-periodic flow with three independent and incommensurate frequencies. This is a
remarkable case since \cite{NRT78} have argued that generic 3-frequency flows are usually
unstable.

\begin{figure}
\centering
\includegraphics[width= 0.75\textwidth,height= 0.5\textwidth]{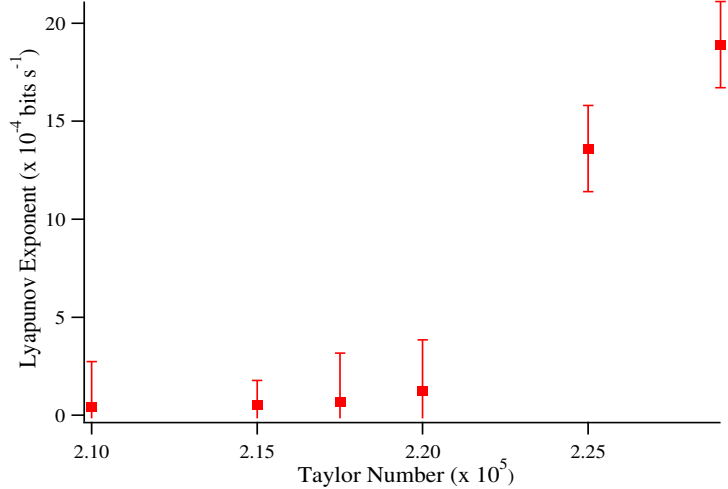}
\caption{\label{Lyapunov}The largest Lyapunov exponent for the 2AV and 2MAV flows against
Taylor number.}
\end{figure}
The Poincar\'e section for the other 2MAV cases at $Ta\geq 225,000$ appear to be more typical
of chaotic flows as illustrated by Figure~\ref{Delay2} (f) for $Ta= 229,250$.
Figure~\ref{Delay2}~(f) only shows the positive crossings for reason of clarity but the
crossings in the negative direction are qualitatively the same and are found in a region which
overlaps with that occupied by the positive crossings.  This is in contrast to the
quasi-periodic 2MAV where the crossing the different directions occupy distinct and
well-separated regions.  As a test for low-dimensional chaotic behaviour, the largest Lyapunov
exponent, $\lambda_1$, was calculated for all the AV and MAV cases, using the method of
\cite{WSSV85}.  As Figure~\ref{Lyapunov} shows, the flows at low Taylor number up to, and
including, the 2MAV case at $Ta= 220,000$ have a largest Lyapunov exponent which cannot be
distinguished from zero. However, $\lambda_1$ is significantly positive for the last two 2MAV
cases.  A positive Lyapunov exponent is clear evidence for chaotic solutions, indicating that
a clear transition to low-dimensional chaotic behaviour occurs, but only \emph{after} the
quasi-periodic transition to 2MAV. The intermediate 2MAV case has a value for $\lambda_1$
which is indistinguishable from zero, confirming the existence (hinted at above) of a
nonchaotic three-frequency flow.



\subsection{Transition to $m = 3$ and higher modes}\label{m3}
As $Ta$ was increased through the 2MAV regime discussed above, the flow dominated by $m = 2$
was found to give way to a new type of flow dominated by $m = 3$ at $Ta = 2.3\times 10^5$. We
consider in this subsection the flow structure and transitions from the $m = 3$ state as
Taylor number is varied.

\subsubsection{Basic flow structures}
\begin{figure}
(a) \hspace{0.48\textwidth}(b)\\
\includegraphics[width=0.47\textwidth, height=0.47\textwidth]{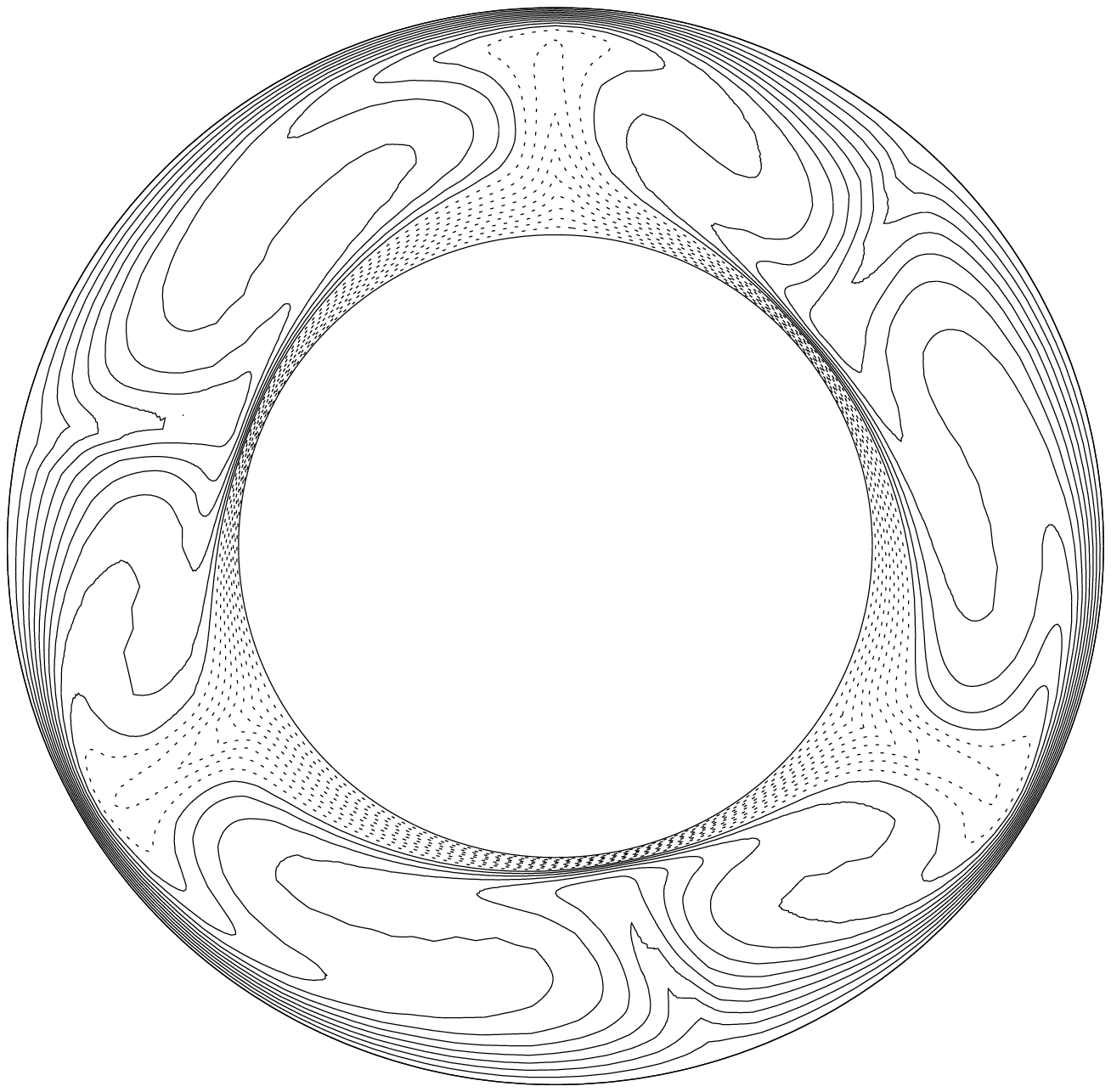}
\hfill
\includegraphics[width=0.47\textwidth, height=0.47\textwidth]{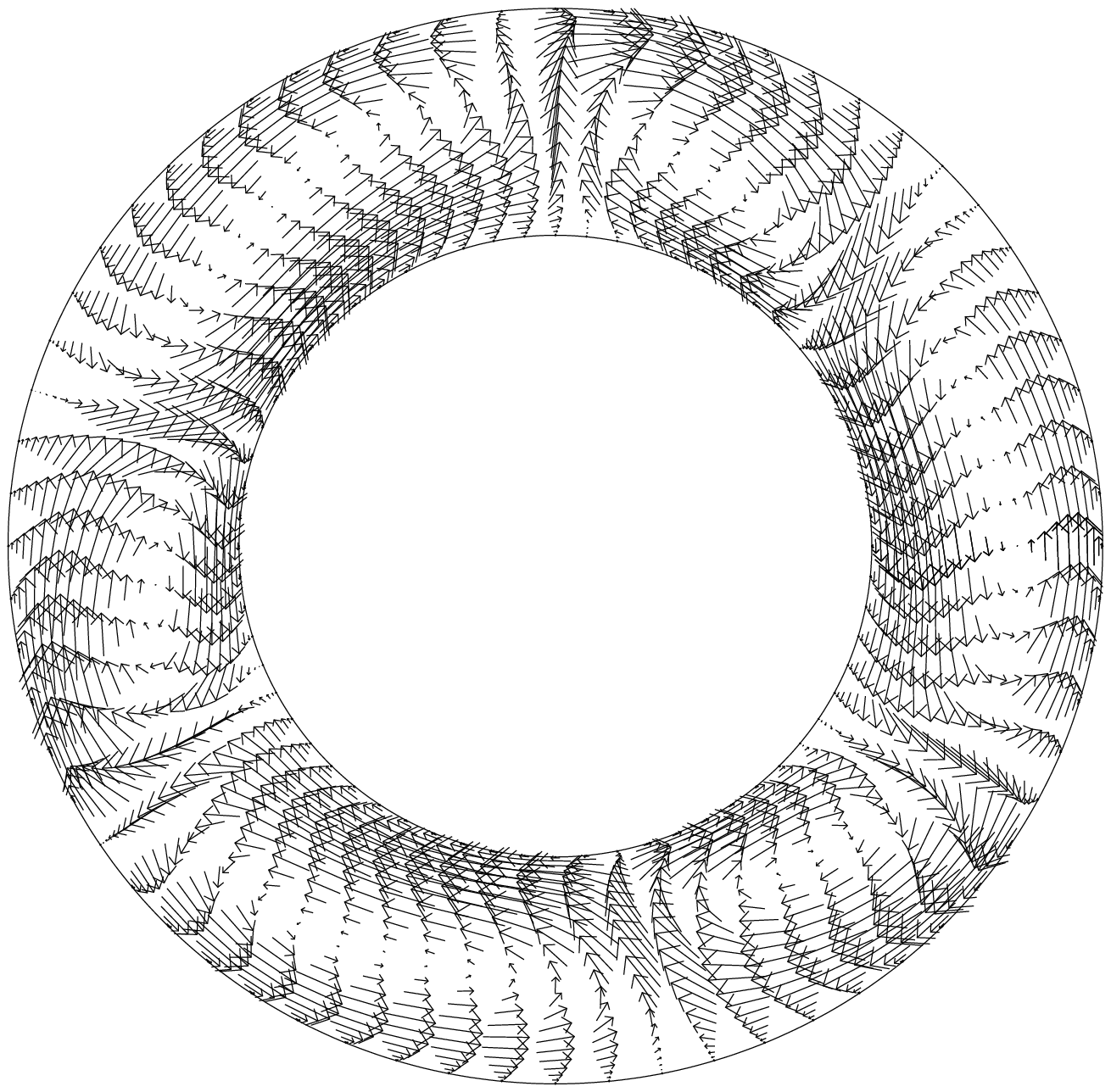}
\caption{\label{Field3} Horizontal cross-sections at mid-height of a wave-3 flow at
$Ta=1.5\times 10^6$. (a) Temperature contours and (b) the horizontal velocity field.}
\end{figure}
\begin{figure}
\includegraphics[width=\textwidth]{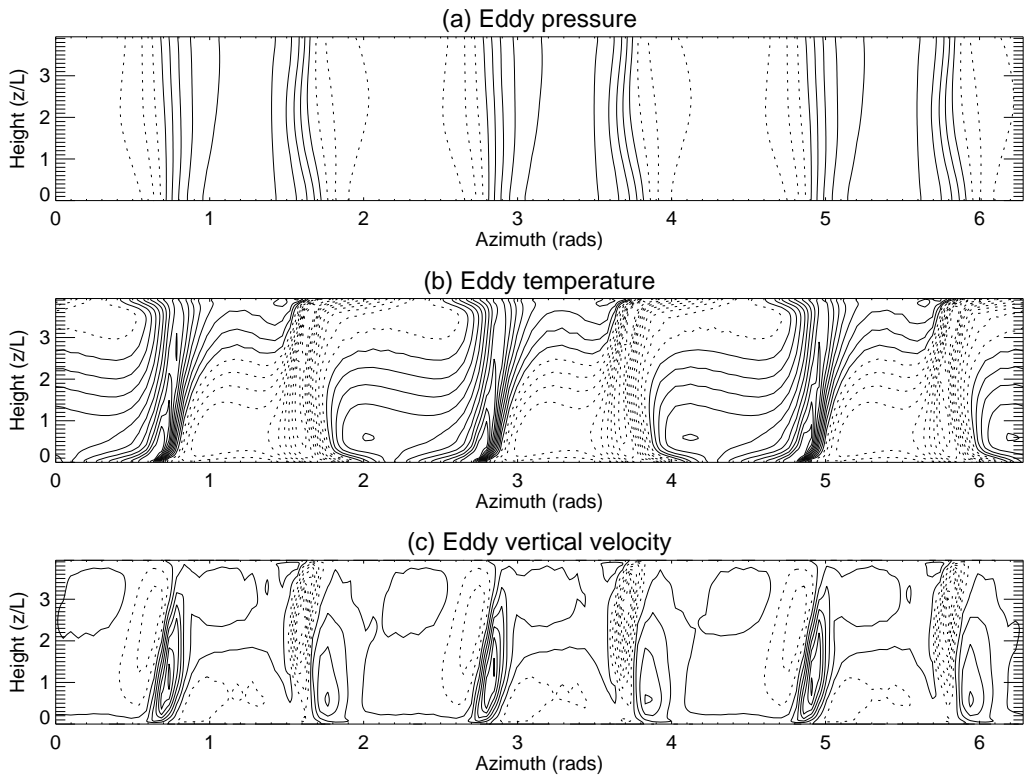}
\caption{\label{TZfields2}Azimuth-height maps of eddy fields (with azimuthal flow removed) in
a typical simulation of the wave-3 flow at $Ta = 1.5 \times 10^6$ using air as the working
fluid. (a) Pressure (contour interval = 0.25 dimensionless units), (b) Temperature $T$
(contour interval = 0.05, normalised to an imposed temperature difference of $\pm 1.0$) and
(c) Vertical velocity (contour interval = 0.5 dimensionless units). }
\end{figure}
\begin{figure}
\includegraphics[width=\textwidth]{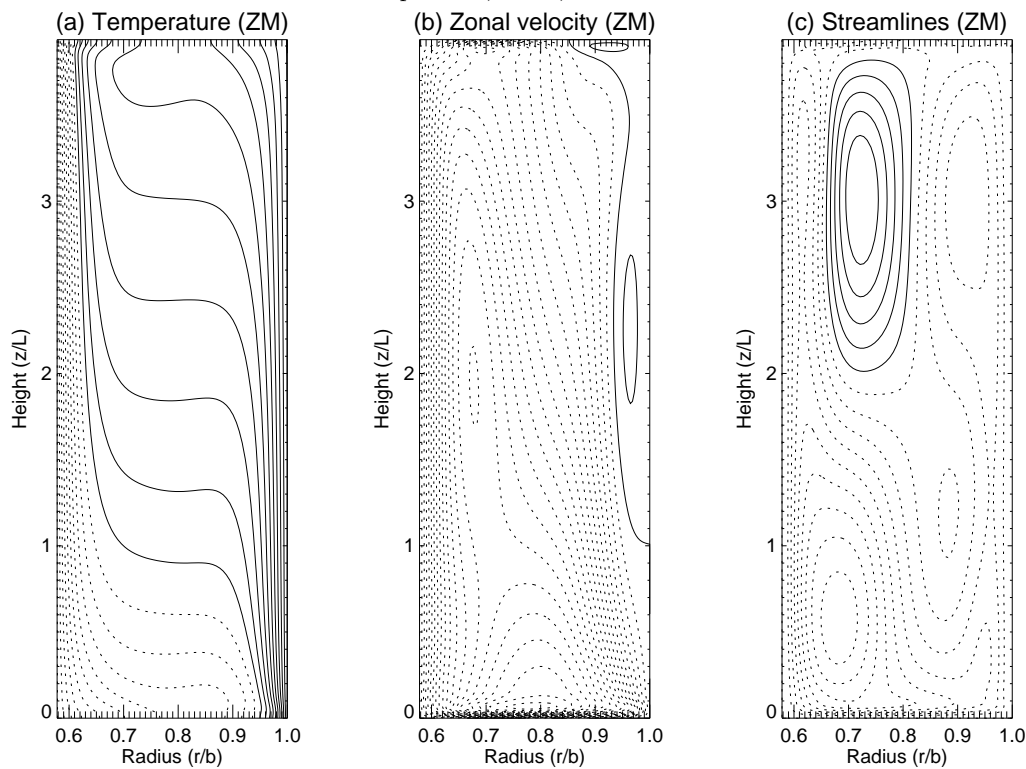}
\caption{\label{ZMfields2}Zonal mean fields for the typical wave-3 simulation at $Ta = 1.5
\times 10^6$ using air as the working fluid. (a) Temperature (contour interval = 0.1,
normalised to an imposed temperature difference of $\pm 1.0$), (b) zonal velocity $V_\phi$
(contour interval = 0.25 dimensionless units) and (c) meridional stream function (contour
interval = 0.01 dimensionless units). }
\end{figure}

The basic flow pattern is illustrated in Figures~\ref{Field3} to \ref{ZMfields2} for a steady
wave, 3S, at $Ta=1.5\times10^6$.  Figure~\ref{Field3} shows maps in the $(r,\theta)$ plane of
temperature and velocity vectors at mid-height for a typical $m = 3$ case. The temperature
fields show the presence of pronounced `plumes' alternately of warm and cool air, highly
concentrated in azimuth and crossing the annular gap radially in association with strong
radial jets.

The barotropic nature of the fully developed waves in this region of parameter space is
illustrated in Fig.~\ref{TZfields2}, which shows azimuth-height sections of an $m = 3$ flow at
$Ta = 1.5 \times 10^6$. The pressure field (Fig.~\ref{TZfields2}(a)) clearly shows very little
phase tilt with height, in complete contrast to those in the 2AV regime shown in
Figure~\ref{TZfields1}. The temperature field in Figure~\ref{TZfields2}(b), too, has a much
more complicated structure than for the $m=2$ flows, with highly concentrated plumes of hot
and cold air which have very little slope with height, interspersed with broad regions with
relatively very weak horizontal thermal gradients. The thermal plumes are aligned azimuthally
with the regions of strong azimuthal pressure gradient, consistent with optimising the
correlation between radial (geostrophic) velocity and temperature perturbations. Similarly,
the regions of strong upward vertical velocity (Fig.~\ref{TZfields2}(c)) are also concentrated
close to the strongest positive temperature anomalies, and vice versa, consistent with an
optimisation of $\overline{w'T'}$ as before. Regions of strong downward motion (necessary to
satisfy mass conservation) are concentrated in plumes or jets adjacent to the strong upwelling
jets, forming `cross-frontal' circulations which show some similarities with those inferred
for atmospheric frontal regions in developing cyclones (\cite{Hos82}). The overall impression
is that the flow in this region of parameter space is strongly nonlinear and much modified
from the simple, linearly unstable Eady solution of the corresponding state for $m = 2$ flows
at much lower Taylor number.

The increasingly barotropic and nonlinear character of the flow at high $Ta$ is also apparent
in the azimuthal mean flow structure, which is illustrated in Fig.~\ref{ZMfields2}. This time,
in contrast to the low $Ta$ states (fig.~\ref{ZMfields1}), the temperature field
(Fig.~\ref{ZMfields2}(a)) exhibits strongly developed sidewall thermal boundary layers and
relatively weak horizontal thermal gradients in the interior. The near-absence of horizontal
thermal contrast in the interior would suggest (via the geostrophic thermal wind relation)
that the zonal flow should be relatively barotropic. This is confirmed in
Fig.~\ref{ZMfields2}(b), which shows negative (retrograde) azimuthal flow at most heights
except near the outer cylinder. The zonal mean meridional streamlines
(Fig.~\ref{ZMfields2}~(c)) exhibit an additional (thermally indirect) 'Ferrel' cell, in
contrast to Fig.~\ref{ZMfields1}~(c), indicating a strong contribution of baroclinic eddies to
the heat transport in this case.

\subsubsection{Sequence of flow transitions}
\begin{figure}
\includegraphics[width=\textwidth]{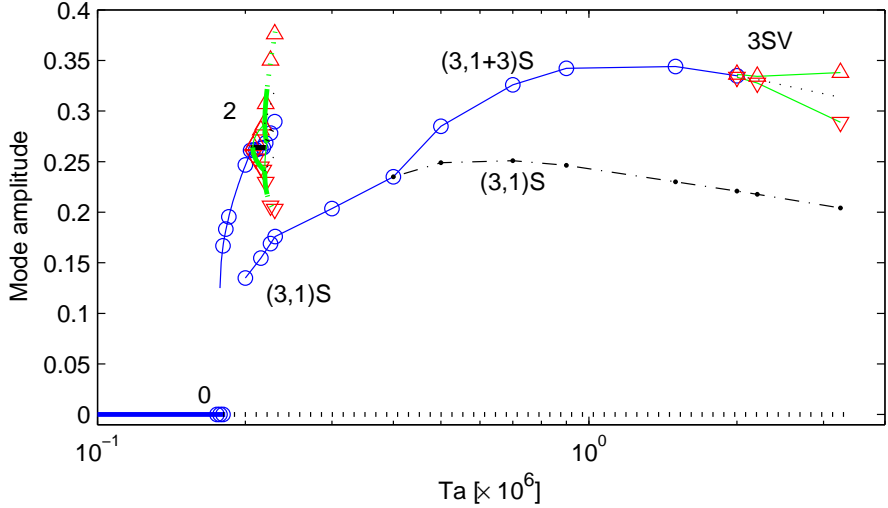}
\caption{\label{Amplitude3} The amplitude of the $m=3$ solutions against the Taylor number.
The solid line and circles show the sum of the first three radial modes of the azimuthal mode
$m=3$, while the dash-dotted line only shows the contribution of the first radial mode.  For
the vacillating flows, the mean amplitude is given by the dotted lines, and the amplitude
extrema are shown by the solid lines and triangles.  To put the $m=3$ solutions into context,
the transition from axisymmetric flow and the $m=2$ solutions, shown in Figure~\ref{Amplitude}
are reproduced.}
\end{figure}

\begin{figure}
\includegraphics[width=\textwidth]{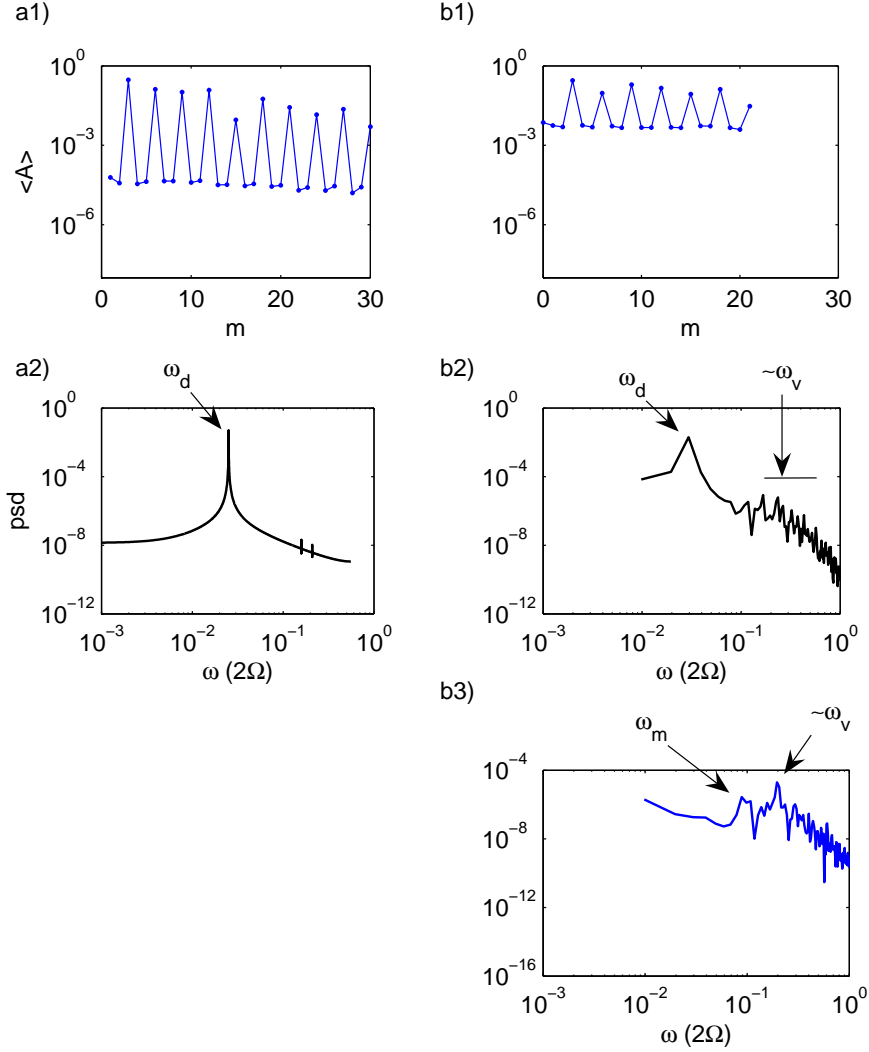}
\caption{\label{Spectra3}  Spectra of representative cases of the $m=3$ flows from time series
of the amplitudes of the azimuthal modes at mid-radius and mid-height. Column a) steady wave
3S, at $Ta= 0.9\times10^6$, and column b) irregular structural vacillation, 3SV, at
$Ta=3.25\times10^6$. Top row 1): spatial spectra of the time-averaged amplitude of the
azimuthal wave mode, $m$. Middle row 2): temporal spectra of the time series of the cosine
component of mode $m=3$;  Bottom row 3): temporal spectra of the amplitude of $m=3$. The main
frequencies are indicated by their annotation.}
\end{figure}

The onset of $m = 3$ effectively began a new sequence of bifurcations, since the $m = 3$ flow
at $Ta=230,000$ converged to a regular pattern of a steady amplitude of $m=3$ and its
harmonics, as shown by the time averaged amplitude spectra in row 1 of Figure~\ref{Spectra3}.
The new solution showed substantial hysteresis relative to the previous $m=2$ regime in that
reducing the Taylor number to $Ta=200,000$ still gave a steady $m=3$ solution. This 3S regime
persisted until $Ta = 1.5 \times 10^6$ but at $Ta=2\times10^6$ other frequencies emerged.  The
sequence of $m=3$ flows observed can be illustrated by Figure~\ref{Amplitude3} which shows the
amplitude of the azimuthal mode $m=3$, where the solution branch of the $m=2$ flows
(Fig.~\ref{Amplitude}) is reproduced to put the $m=3$ observation in context.

The sum of all radial modes of the azimuthal mode at mid-radius resulted in an initial
increase of the amplitude on increase of the Taylor number followed by a decrease, which was
mirrored by a continuing increase of the amplitude at a radius of one quarter of the gap from
the inner wall three quarters.  This suggested that the radial structure of the flow was more
complex. Approximating the radial structure by half sine waves, the approximate distribution
into the first three radial modes, $A_{m,n}$ with $m=3$ and $n= 1, 2, 3$, can be estimated
from the amplitudes, $A_{1/4}, A_{1/2}, A_{3/4}$ at a radial position of a quarter, a half,
and three-quarters of the gap, respectively, by
\begin{eqnarray*}
n=1 & : & A_{3,1}= \frac{1}{\sqrt{8}}\left(A_{1/4} + A_{3/4}\right) + \frac{1}{2} A_{1/2}  \\
n=2 & : & A_{3,2}= \frac{1}{2} \left(A_{1/4} - A_{3/4}\right) \\
n=3 & : & A_{3,3}= \frac{1}{\sqrt{8}}\left(A_{1/4} + A_{3/4}\right) - \frac{1}{2} A_{1/2}  .\\
\end{eqnarray*}
Figure~\ref{Amplitude3} then shows the amplitude of the sum of first three radial modes as the
solid line, labeled with '(3,1+3)S' since the second radial mode did not contain any
substantial amplitude.  Below $Ta\approx 400,000$, the overall amplitude is fully described by
the first radial mode only.  Following this branch to lower values of the Taylor number, one
can observe a substantial regions of hysteresis between the $m=2$ branch and the $m=3$ branch.
Between $Ta= 400,000$ and $Ta=500,000$, the radial structure of the flow changes whereby
$(m=3,n=3)$ grows beside $(m=3,n=1)$.  In terms of the horizontal structure of the flow, this
indicates the emergence of a flow with wave trains near either side wall.  Finally, this
solution develops a vacillation which appears as a radial fluctuation of the eddies or wave
maxima.  Visually, this vacillation would be classified as a structural vacillation since the
overall amplitude (still dominated by the first radial mode) varies little, but the exchange
between flow or temperature maxima at mid-radius and near the walls varies the structure of
the flow.

\subsubsection{Spatial and temporal spectra}
Figure~\ref{Spectra3} shows a set of time-averaged mode spectra in row (1) and corresponding
power spectra for the cosine component of $m=3$ (row 2) and for the amplitude of $m=3$ in row
(3). A steady wave case and a vacillating case are shown. While the transition from 2S to 2AV
discussed above was characterised by a fundamental change in the time-averaged spectrum, the
mode spectra for all wave-3 flows are qualitatively very similar. The only substantial change
is the gradual increase of the 'floor' from around $10^{-5}$ to around $10^{-3}$.

Because the vacillation of the amplitude is very weak in almost all cases, the power spectrum
of the cosine component of $m=3$, the upper of the spectral curves in each plot in row 2 of
Figure~\ref{Spectra3}, is very bland and only shows the distinctly the peak at the drift
frequency of the wave. The power spectrum of the amplitude, however, shows two broad but
distinct peaks, one associated with vacillation of the flow structure and the other with a
slower modulation.  We did not observe any periodic vacillation in this set of simulations. It
should be noted, however, that the transients leading to the steady 3S solutions were
characterised by a decaying, periodic oscillation with oscillation frequencies comparable to
the vacillation frequency identified in Figure~\ref{Spectra3}~(b3).

\subsubsection{Phase portraits}
\begin{figure}
(a)\hspace{0.48\textwidth}(b)\\
\includegraphics[width=0.49\textwidth]{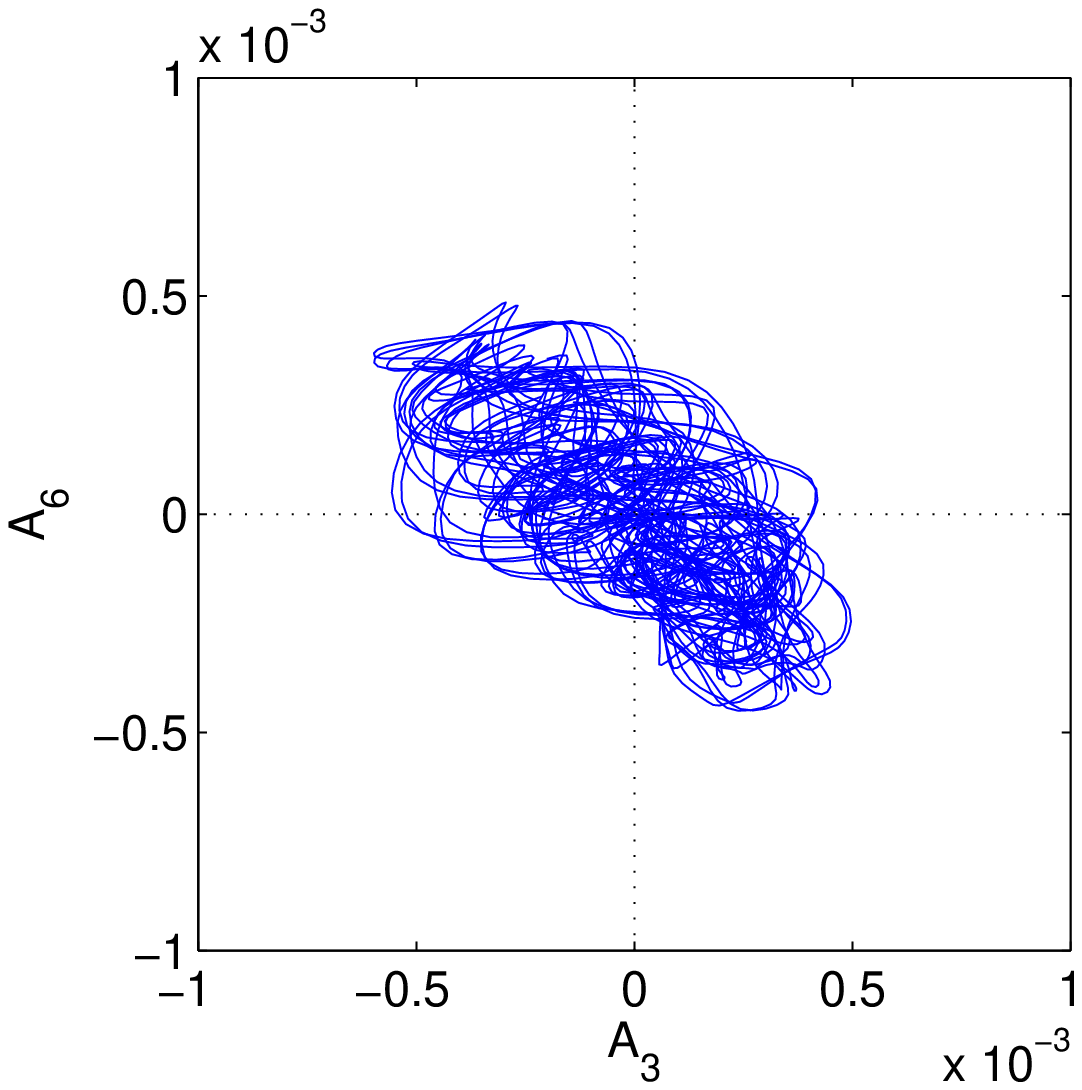}
\hfill
\includegraphics[width=0.49\textwidth]{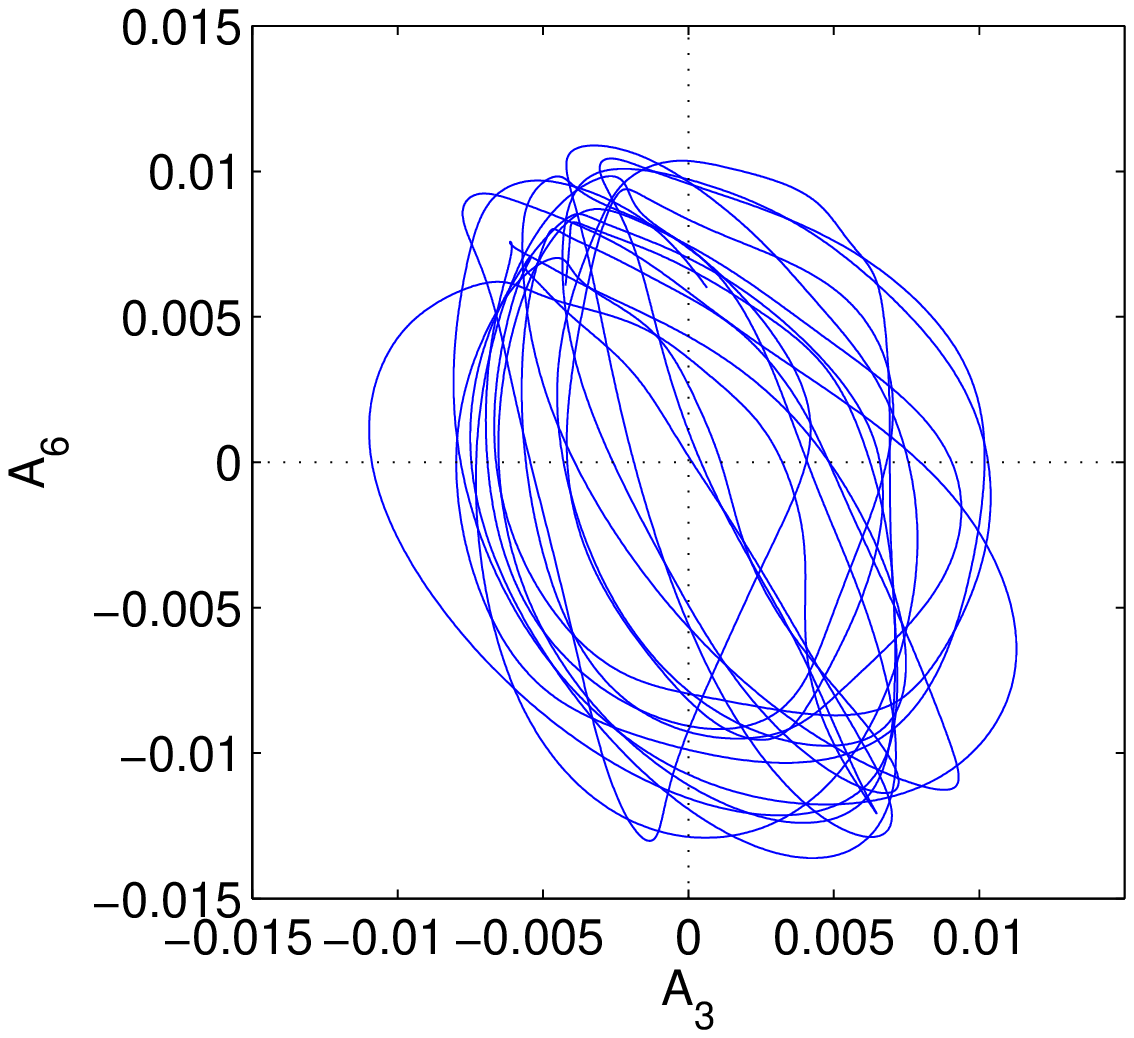}\\
(c)\hspace{0.4\textwidth}(d)\\
\includegraphics[width=0.49\textwidth]{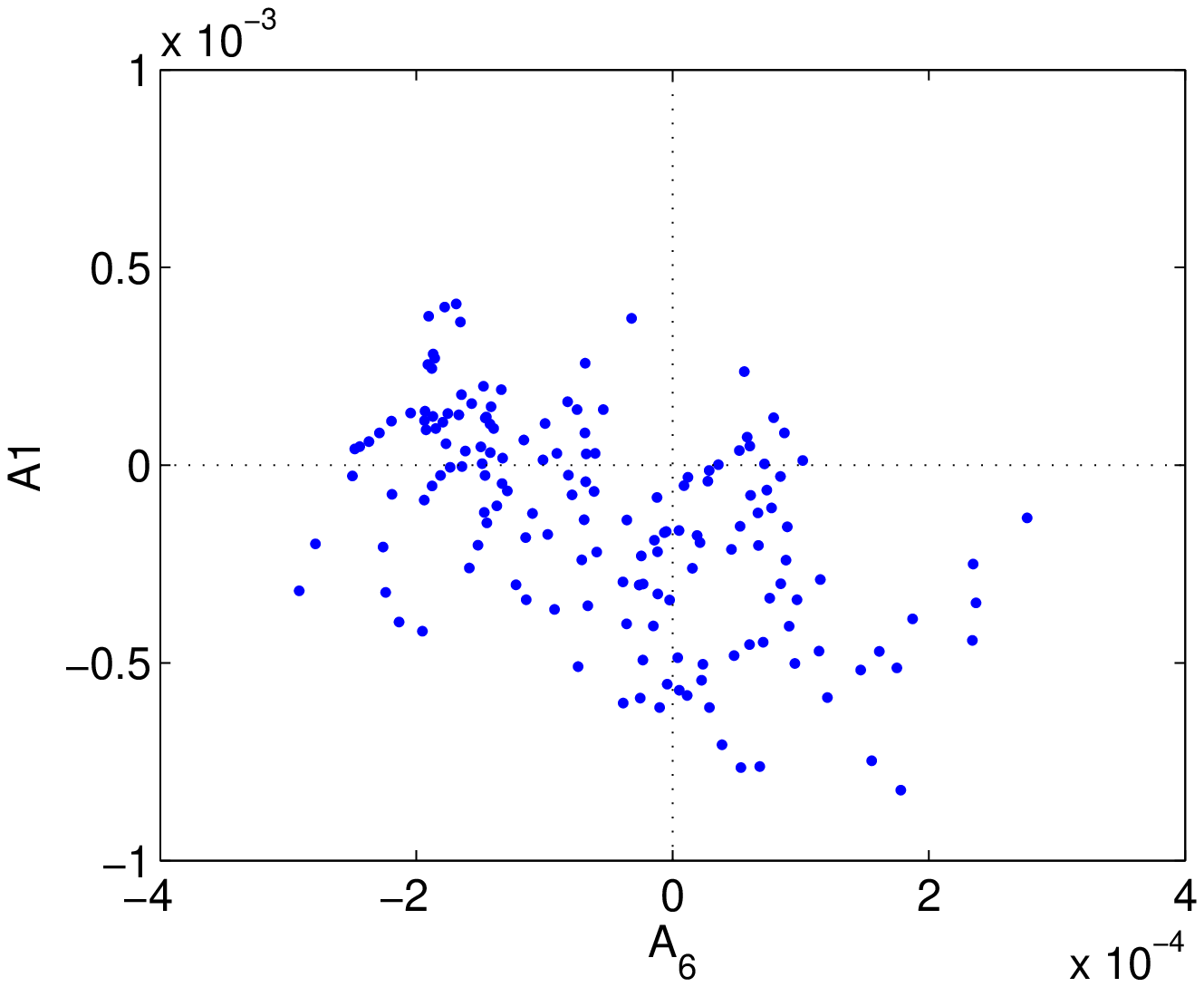}
\hfill
\includegraphics[width=0.49\textwidth]{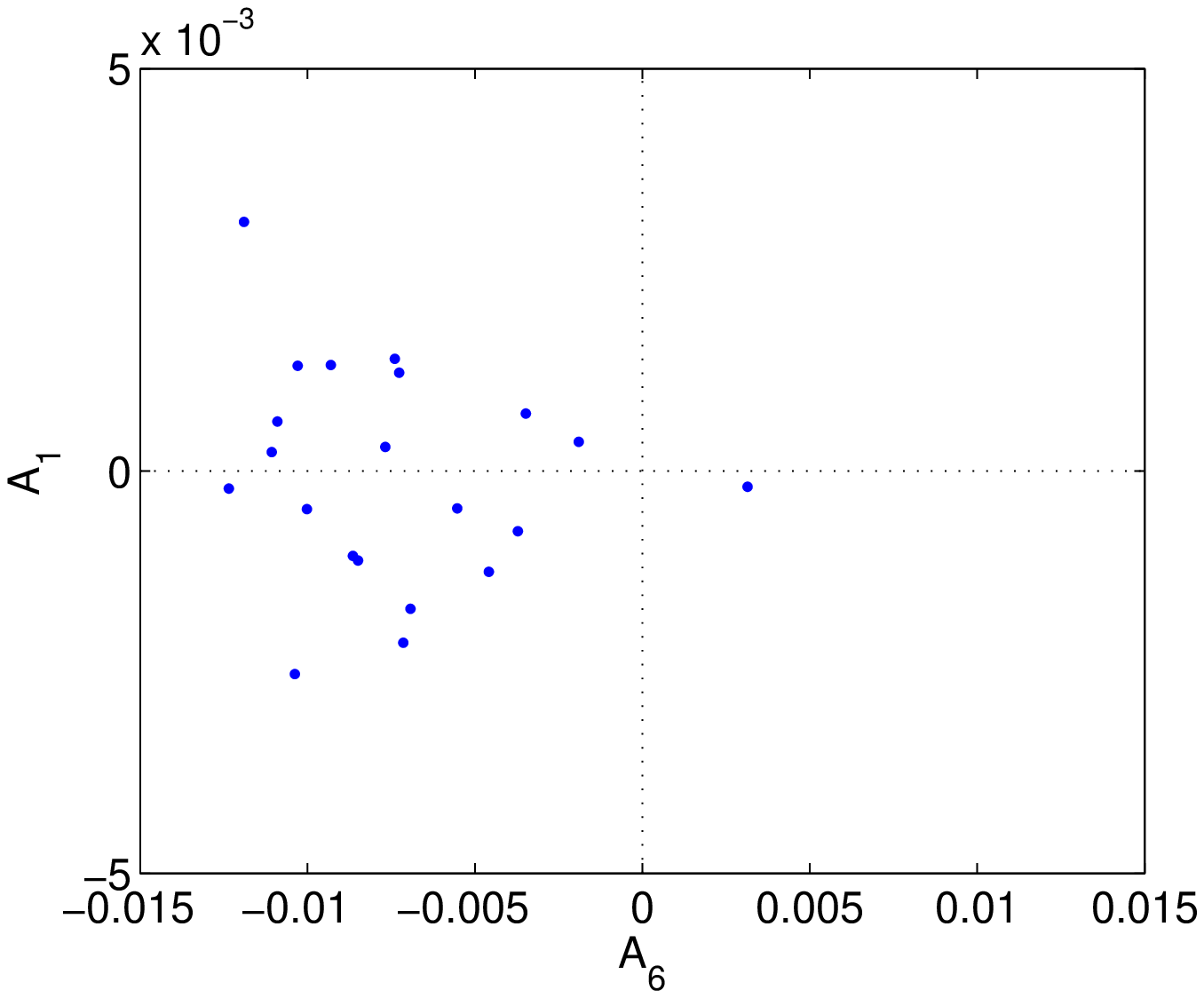}\\
\caption{\label{phase1} Phase portraits and Poincar\'e sections in the $m = 3$ regime: (a)
Phase portrait of the amplitude of $A_6$ against $A_3$ for $Ta = 2.0 \times 10^6$, (b) Phase
portrait for $Ta = 3.25 \times 10^6$; (c) Poincar\'e section at $A_3=0$ showing $A_1$ against
$A_6$ for $Ta=2.0\times10^6$.  (d) Poincar\'e section as (c) but for $Ta=3.25\times10^6$. }
\end{figure}
Figure~\ref{phase1} shows phase portraits and Poincar\'e sections of two vacillating cases,
the first time-dependent case at $Ta=2\times10^6$ and the final case at $Ta=3.25\times10^6$.
Since the strength of the vacillation is very small compared to the mean amplitude, only the
phase portraits generated from the wave amplitudes are shown; the phase portraits for the
cosine or sine components would only show the circle representing the wave drift.  Both cases
show irregular oscillations of the amplitude in the phase portrait of the amplitude of mode
$m=6$ against that of $m=3$.  The Poincar\'e section of the phase portrait for the lower
Taylor number in Figure~\ref{phase1} (c), which shows the amplitude of $m=1$ against that of
$m=6$ at the section where $A_3=0$ reveals no obvious structure at all.  The corresponding
section for the high Taylor number is also irregular but it appears that at least the negative
crossing points are arranged around a circular structure indicating the possibility of a fuzzy
torus.  It has to be noted that the scale of the plots varies substantially between the two
cases with the $y$-axis 5 times larger and the $x$-axis 50 times larger in Figure~\ref{phase1}
(d) compared to Figure~\ref{phase1} (c). Due to the computational expense of the calculations
at these high Taylor numbers, the time series at the highest value of Ta, only covers a few
drift periods of the dominant flow feature which is too short to estimate Lyapunov exponents.
For the same reason, it has not been possible to carry out more integrations between
$Ta=2.2\times10^6$, which is qualitatively and quantitatively similar to $Ta=2.0\times 10^6$,
and $Ta=3.25\times10^6$ to investigate whether the toroidal structure seen in
Figure~\ref{phase1} (d) is a larger-amplitude version of the fluctuations seen in
Figure~\ref{phase1} (c) or whether it emerges as a new type of oscillation besides the
small-scale fluctuations in a flow regime transition.  As a final note it should be mentioned
that no quasi-periodic solutions were found in the simulations carried out.

\section{Discussion}
\subsection{Regime diagram and onset of baroclinic waves}
The overall structure of the regime diagram, presented in Figure~\ref{Regime}, is
qualitatively and quantitatively consistent with previous experiments and simulations of the
baroclinic annulus using liquids instead of air.  They all show the characteristic anvil
shaped transition curve between axisymmetric flow and regular waves, separating the regimes
into a wave regime, an upper symmetric regime at large $\Theta$, and a lower symmetric regime
at small $\Theta$/small Taylor number, e.g.~\cite{Fein73,FP76,hide58,HM75} and \cite{J81}. The
upper symmetric regime indicates a flow at large $\Theta$ or thermal Rossby number which is
stabilised by strong stratification.  The transition from the upper symmetric regime to
baroclinic waves appears largely controlled by a critical thermal Rossby number.  The lower
symmetric regime is stabilised by viscous dissipation, and the transition to waves is largely
given by a critical Grashof number or Rayleigh number.

Most of these had been performed with fluids of Prandtl number $Pr>\sim7$ but even an early
study by \cite{FP76} using mercury, with $Pr=0.025$, showed the same qualitative
characteristic anvil shaped stability curve although the onset of waves was shifted towards
higher Taylor numbers by several orders of magnitude.  The case of air with a Prandtl number
between that of mercury and water, however, seemed not to be located between those cases but,
if anything, at slightly small values of the Taylor number.  The observations from our
simulations suggest that the case of $\Delta T=30K$ is positioned near the 'knee' of the
classical anvil shape.

The, albeit narrow, hysteresis at the transition between the axisymmetric flow and the steady
wave suggests a subcritical bifurcation.  While \cite{HM78} reported that the transition from
the symmetric regimes in an annulus showed hysteresis only if the upper boundary was a free
surface, \cite{KoWh81} presented evidence for the possibility of small hysteresis for the
transition to or from the upper symmetric regime.  In their experiments, performed with water
($Pr\sim7$), the hysteresis set in at a critical Taylor number below which the transition was
not hysteretic and above which the extent of the overlap of the symmetric and wave regimes
increased with the Taylor number. Hysteresis near the 'knee' of the anvil-shaped marginal
stability curve has not been documented before. Other studies have either reported or assumed
a Hopf bifurcation, e.g.~\cite{Rand82,MB91,LN04}, or the occurrence of so-called \emph{weak
waves} prior to the onset of fully-developed baroclinic waves, \cite{HM78,J80}. The weak waves
have been suggested to be wave modes which are located near the upper or lower boundary and
are decaying in the fluid interior.  No such weak waves were observed in this study, though it
has to be said that weak waves seem to have been observed predominantly at the transition from
the upper symmetric regime at somewhat higher Taylor numbers, well past the knee of the anvil.

\subsection{A bifurcation scenario}
\begin{figure}
\includegraphics[width=\textwidth]{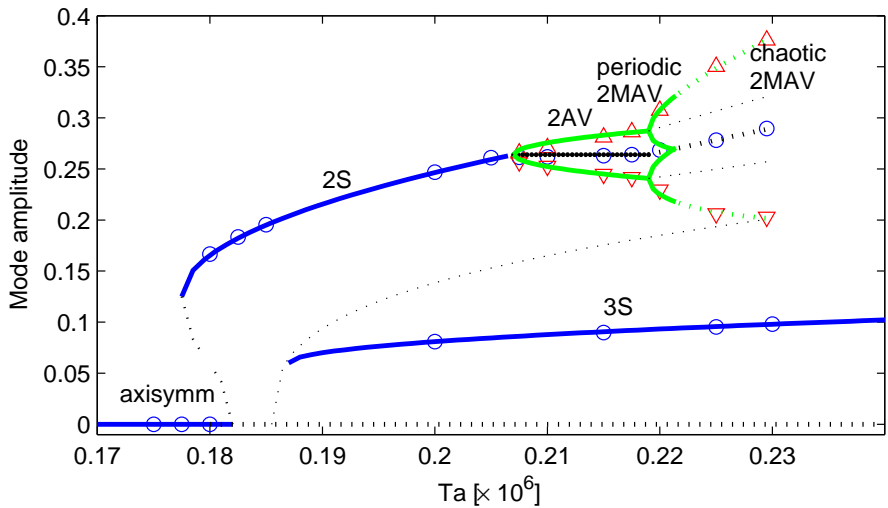}
\caption{\label{Hopf1}The bifurcation scenario for the transition from the upper symmetric
regime to a steady wave and subsequent vacillations for the $m=2$ flows.}
\end{figure}
The transition sequence from the axisymmetric flow through all observed wave 2 flows follows
clear steps of increasing complexity, from axisymmetric flow via a steady wave with one
frequency, a two-frequency flow of a vacillating wave, a quasi-periodic three-frequency flow
of a modulated vacillating wave to, finally, a chaotic flow.  As Hopf bifurcations for the
onset of baroclinic waves and the subsequent onset of amplitude vacillation are often cited,
e.g.~\cite{Rand82,MB91,LN04},  it seemed obvious to fit the normal form of a Hopf bifurcation
to the wave amplitude data. This transition scenario is reproduced in Figure~\ref{Hopf1} by
showing the mean amplitude and the envelope of the vacillation.

Since the initial onset of waves is a subcritical transition, a subcritical Hopf bifurcation
and a Fold or saddle-node bifurcation is invoked, e.g.~\cite{GH83,Rand82}.  A linear
regression to a curve of the form $A=A_0 + [(Ta-Ta_S)/q]^{1/2}$, using the five data points
from the 2S regime, gave $A_0=0.125$ for the wave amplitude at the saddle-node, a growth
parameter of $q=1.50\times10^{6}$, and a critical Taylor number for the saddle-node of
$Ta_S=177,500$ with a correlation coefficient of $r^2=0.9997$. The 'unstable branch' has been
filled in using this critical Taylor number and a Taylor number for the location of the
unstable Hopf as $Ta_0= 182,000$.

The following transition, from the steady wave to the periodic oscillation of the wave
amplitude in the 2AV flow, is consistent with a second Hopf, also known as a Neimark
bifurcation.  This is indicated by fitting a square-root function to the vacillation strength
(amplitude maximum minus mean amplitude) for the four simulations in the 2AV regime as a
function of the Taylor number. While this is a standard bifurcation, it is in contrast to all
experience for the onset of amplitude vacillation in a liquid with a range of Prandtl number
between 7 and 30, where the evidence is that vacillation occurs on decreasing the Taylor
number from the steady wave. In those cases, the vacillation is a precursor to a particular
wave mode giving way to, and co-existing with, a flow of lower mode number. The experiments in
mercury by Fein and Pfeffer from 1976 at much lower Prandtl number did not report such details
in the wave regime. The consensus for the other experiments in liquids is that vacillation is
much more common in higher Prandtl number, where most of the wave 2 and 3 domains at $Pr=26$
in \cite{FR97} were time-dependent, whereas at $Pr=13$ and $Pr=7$ substantial ranges were
found where a steady wave was reported by \cite{Hignett85,Hetal85,JJF81}. Some initial
experiments carried out in the apparatus used by \cite{HM78,RBJS92,FR97} but now using air
confirmed at least qualitatively the onset of vacillation with increasing Taylor number,
consistent with these computational results.

The third transition, to a modulated 2MAV was a not surprising continuation of the
'quasi-periodic route to chaos' described by \cite{NRT78}, see also \S6 in \cite{Sc95}, but
the nature of the initial solution as a quasi-periodic 2MAV with three independent frequencies
was unusual. As was shown by \cite{NRT78}, generic three-frequency flows are expected to be
chaotic rather than periodic.  Previous illustrations of the existence of three-frequency
quasi-periodic orbits were demonstrated numerically, with the modulation of a simple circle
map by \cite{GrOtYo83}, and in experiments of convection of mercury in a magnetic field by
\cite{LF+83}.  However, no previous example such a flow has been reported from either
numerical or experimental studies of baroclinic waves.  All previous reports of quasi-periodic
modulated vacillations, e.g.~by~\cite{RBJS92} and \cite{FR97}, involved frequency-locking
between the drift frequency and the vacillation or modulation frequency. Conversely, no
instance of frequency-locking was observed in the current numerical simulations. This supports
the previously suggested notion that the ubiquitous frequency locking in the baroclinic
annulus relies on a coupling between the spatial phase of the wave and its amplitude by
imperfections in the apparatus or the presence of sensors in the fluid.

The final type of flow dominated by $m=2$ was a chaotic 2MAV.  In the average statistics, such
as the time-averaged wave spectrum and the average drift frequencies, there was little
difference between the quasi-periodic and the chaotic versions of the 2MAV\@. Fitting yet
another square-root function to the modulation amplitude over the vacillation amplitude in the
MAV regimes suggests that the onset of modulation in itself could be consistent with a further
supercritical Hopf bifurcation.  As there are only three sufficient 2MAV simulations, together
with a change in character from quasi-periodic to chaotic, the assumption of a square-root
scaling is conjecture but gives a reasonable representation of the development of the
modulation. Square-root dependent or not, the growth of the modulation hints at the possible
mechanism for the transition from periodic to chaotic behaviour: the minimum of the amplitude
maximum and the maximum of the amplitude minimum intersect in Figure~\ref{Hopf1} at a point
between the observed periodic flow at $Ta=220,000$ and the first chaotic flow at $Ta=
225,000$.  It appears therefore, that the transition to chaos in this example might not have
arisen from the local stretching of the torus, \S6.1 in \cite{Sc95}, but from a crisis as
suggested by ~\cite{GO+83} where the orbit of the quasi-periodic 2MAV intersects the unstable
orbit of the 2AV, see also \S6.4 in \cite{Sc95}.

The final transition from the $m=2$ branch to the steady 3S on the $m=3$ solution branch could
also be due to a crisis, this time caused by the collision of the chaotic 2MAV attractor with
the unstable orbit separating the $m=2$ and $m=3$ solutions. Such a collision is possible in
phase space because the 2MAV flow involves finite-amplitude variations of the $m=3$ mode, and
it is consistent with the conjecture by \cite{GO+83} that 'almost all' sudden changes in
chaotic attractors are due to crises.  The existence of an unstable orbit separating the basin
of attraction of the $m=2$ and $m=3$ solutions is indicated by the dotted line between the two
solution branches, following Figure 8 of \cite{LeNa03}.

\subsection{Transitions on the $m=3$ branch}
The $m=3$ flows also showed a progression from periodic flow of a steady wave, 3S, to
irregular vacillations but following a very different route.  This route involved a change
in the radial structure of the flow while remaining a simply-periodic flow of a steady wave.
No quasi-periodic flows were observed but the possibility remains that those could exist at
Taylor numbers not covered by the simulations.

The vacillating flows at the very high Taylor number are qualitatively similar to experimental
evidence by \cite{FR97} at $Pr=26$ which they classified as 'Structural Vacillation', also
known as 'Tilted-trough vacillation'.  That flow was largely dominated by a simple azimuthal
mode ($m=2$ in their case), superimposed on which were small fluctuations consistent with a
higher radial mode. While the large scale structure of the flow was highly regular, the
fluctuations were shown by them to be not consistent with low-dimensional dynamics. The
failure to compute reliable dimension estimators for structural vacillation had previously
been reported in two independent experimental studies by \cite{RBJS92} and \cite{GB83}.

The significantly different vertical structure of the steady 3S, by being much more
barotropic, could result in a different type of instability altogether.  The radial profile of
the azimuthal velocity, shown in Figure~\ref{ZMfields2} (b) suggests a Reynolds number of
$Re\equiv UL/\nu\approx50$ for the boundary layer near the inner wall (using a velocity of
$\sim12U$ at a distance of $L\sim0.1b$ from the wall).  Studies of the instability of detached
shear layers by \cite{FR99} suggest that such a Reynolds number is sufficient for the onset of
barotropic instability in a Hopf bifurcation which leads to a string of vortices along the
shear layer with a wave number largely determined by the thickness of that layer. With a
critical Taylor number of $Ta_c= 428,000$, the observed growth of the higher radial mode,
shown in Figure~\ref{Amplitude3}, is consistent with a square-root dependence up to the Taylor
number of $Ta=900,000$.  Barotropic instability has also been proposed by \cite{SchD90} as a
mechanism active in the decay of baroclinic waves in the atmosphere. If this instability has
caused the onset of the waves near the side walls, it is not surprising that the
quasi-periodic route to chaos was not followed, as the experiments by \cite{FR99} did not
observe quasi-periodic vacillating flows either. The indication of steps of a period-doubling
cascade were observed at large Rossby numbers, whereas the increase in Taylor number in the
baroclinic annulus would result in reducing the Ekman number in their study.  At a Reynolds
$Re=O(50)$, they observed only steady and noise waves of wave number $m\geq6$ and a regime of
weak irregular fluctuations (cf their regime diagram in the $Re-E$ plane in Figures 3 in
\cite{FrNi03}). \cite{FrNi03} suggested that the onset of time-dependence in the amplitude of
barotropic vortices may rely on local, small-scale vorticity generation at the side walls
rather than a global-mode instability of the flow in the fluid interior.

\section{Conclusions}
\begin{table}
 \centering
\begin{tabular}{lccl}
 & US & & axisymmetric\\
 Subcritical Hopf & $\leftarrow | \rightarrow$ & 2S & baroclinic instability, \\
  with saddle-node &&& zonal symmetry breaking\\
 Hopf & $\longleftrightarrow$ & 2AV & temporal symmetry breaking\\
 Hopf & $\longleftrightarrow$ & 2MAV & three-frequency quasi-periodic\\
 Crisis & $\longleftrightarrow$ & 2MAV & chaos\\
 Crisis & $\leftarrow | \rightarrow$ & 3S & mode transition, \\
 &&& zonal symmetry breaking\\
 Pitchfork& $\longleftrightarrow$ & (3,1 + 3,3)S & barotropic instability, \\
 &&& radial symmetry breaking\\
 ??? & $\longleftrightarrow$ & 3SV & irregular fluctuations\\
\hline \vspace*{0.5ex}
\end{tabular}
\caption{Tentatively suggested bifurcation sequences from the initial instability to the onset
of the irregular vacillation of the $m=3$ flow. $\longleftrightarrow$ denote transitions
without hysteresis (supercritical bifurcations), and $\leftarrow | \rightarrow$ indicate the
presence of hysteresis. }
 \label{Bifurcations3}
 \end{table}
We have presented results from direct numerical simulations of convection in a baroclinic
annulus filled with air and put those in the context of sloping convection in liquid-filled
baroclinic annulus experiments.  This investigation, and the two previous studies by Maubert
\& Randriamampianina (2002,2003) leading up to this work, followed the most common procedure
of keeping the temperature difference fixed while varying the rotation rate of the system.
Maubert \& Randriamampianina (2002,2003) verified the transition curve between the symmetric
regimes and wave regimes and they reported the main types of baroclinic waves usually observed
in the baroclinic annulus.  The transition curve from the axisymmetric solution to steady,
fully-developed baroclinic waves formed the typical anvil shape in the $Ta-\Theta$ parameter
plane, in common with all published information on this transition for liquids covering a
range of Prandtl number from 0.025 to 26 (though not the extreme case published by \cite{FP76}
for $Pr=63$).

Table~\ref{Bifurcations3} summarises the sequence of flows and bifurcations. The initial
baroclinic instability and the development of the $m=2$ flows are consistent with the standard
quasi-periodic route to chaos with a few individual aspects, such as the subcritical character
of the first Hopf bifurcation and the existence of a rather rare three-frequency
quasi-periodic flow which became chaotic in what could be a crisis.  The mode transition
between flows dominated by $m=2$ and $m=3$, respectively, is also consistent with all previous
evidence for baroclinic waves in the rotating annulus in that it is a sudden mode transition
to a higher wave number on increase of the Taylor number, associated with a substantial
hysteresis.

The results at the highest Taylor number achieved in the study points to a first direct
numerical simulation of a structural vacillation.  The onset and characteristics of structural
vacillation is less well documented in the literature, where a variety of small-scale
processes have been proposed, e.g.~\cite{FR97}, such as inertia-gravity waves observed by
\cite{LoReRi00}.  The evidence here points to the occurrence of a barotropic instability of
the steady wave prior to the onset of temporal fluctuations near the side walls, in which case
vorticity generation at the side walls may be a mechanism in the onset of structural
vacillation in the annulus.

Overall, the observations are consistent with the extensive literature on baroclinic waves in
the thermally forced annulus filled with liquids.  This demonstrates the validity of the
numerical model in parameter ranges not previously covered by direct numerical simulation. The
change in Prandtl number from $Pr\gg1$ to $Pr<\sim1$, however, resulted in the reversal of the
Hopf bifurcation leading to amplitude vacillation. From a physical point of view it seems
sensible to suggest that the relative thickness of the thermal and velocity boundary layers
controls whether the vacillation appears on increase or decrease of the Taylor number. While
this change in itself is rather subtle, it has consequences for the overall organisation of
the regime diagram and the transition from highly regular to complex flows. This point is not
only relevant to baroclinic annulus flows but to all convective flows, if not fluid dynamical
systems in general, where a change in the fluid properties may lead to apparently minor
changes which then cause a fundamental change in the dynamics over a wide range of forcing
parameters.

\acknowledgments We are grateful to the Royal Society for their support of this work
by funding the collaboration between the French and UK partners in a Joint Project
Grant, grant number 15118\@.  The CPU time for this work was provided by the IDRIS
(CNRS, Orsay, France) on their NEC SX-5.


\end{document}